\documentclass[useAMS,usenatbib,epsfig,epsf]{mnras}
\usepackage{times,epsfig}
\usepackage{natbib}
\usepackage{pdflscape,hyperref}  % Landscape pages
\usepackage[T1]{fontenc}
\usepackage{ae,aecompl}
\newcommand{\eone}{EPIC\,215776487}
\newcommand{\etwo}{EPIC\,217280630}
\newcommand{\ethree}{EPIC\,218366972}
\newcommand{\efour}{EPIC\,218717602}
\title[K2 Campaign 7 sdB pulsators]
{Pulsating subdwarf B stars observed with K2 during Campaign 7 and an
 examination of seismic group-properties.}

\author[M.\,D.\,Reed et al.]{
M.\,D.\,Reed$^{1,2}$\thanks{E-mail:MikeReed@missouristate.edu}, 
A.\,Slayton$^1$, A.\,S.\,Baran$^{1,2,3}$, J.\,H.\,Telting$^{2,4,5}$, 
R.\,H.\,\O stensen$^{1,2,6}$,
\newauthor C.\,S\,Jeffery$^7$, M. Uzundag$^{8,9}$, S.\,Sanjayan$^{2,10}$
\\
$^1$Department of Physics, Astronomy and Materials Science,
 Missouri State University, 901 S. National, Springfield, MO 65897, USA \\
$^2$ARDASTELLA Research Group, Institute of Physics, Pedagogical University of Cracow, ul. Podchor\c{a}\.{z}y
ch 2,30-084 Krak\'{o}w, Poland\\
$^3$ Embry-Riddle Aeronautical University, Department of Physical Science, Daytona Beach, FL 32114, USA\\
$^4$Nordic Optical Telescope, Rambla Jos\'e Ana Fern\'andez P\'erez 7, 38711 
Bre\~na Baja, Spain\\
$^5$ Department of Physics and Astronomy, Aarhus University, Ny Munkegade 120, DK-8000 Aarhus C, Denmark\\
$^6$Recogito AS, Storgaten 72, N-8200 Fauske, Norway\\
$^7$Armagh Observatory and Planetarium, College Hill, Armagh BT61 9DG, N. Ireland\\
$^8$ Instituto de F\'isica y Astronom\'ia, Universidad de Valpara\'iso, Gran Breta\~na 1111, Playa Ancha, Valpara\'iso 2360102, Chile\\
$^9$European Southern Observatory, Alonso de Cordova 3107, Santiago, Chile\\
$^{10}$Centrum Astronomiczne im. Miko{\l}aja Kopernika, Polskiej Akademii Nauk, ul. Bartycka 18, 00-716 Warszawa, Polska
}

\date{Accepted
      Received }

\begin{document}

\maketitle

\begin{abstract}
We report the discovery of four new pulsating subdwarf B (sdBV) stars from Campaign 7 of the
Kepler spacecraft's K2 mission. EPICs 215776487, 217280630, 218366972, and 218717602 are all gravity 
(g)-mode pulsators and we also detect two pressure (p)-mode pulsations in \efour .
We detect asymptotic $\ell\,=\,1$ sequences in all four stars, allowing us to identify nearly all of
the g modes. We detect evenly-spaced frequency multiplets in \efour\, from which we determine a rotation
period near seven days. Spectroscopic observations determine that
\ethree\, is in a 5.92\,d binary with most likely a white dwarf companion of canonical mass while the others
have no detected companions. As we detect no multiplets in \ethree , it is added to the growing list
of subsynchronously rotating stars. With 40 Kepler-detected sdBV stars and a growing number of TESS
publications, we update an examination of the group properties to provide direction for models.  We
notice a correlation between effective temperature and period of maximum pulsation amplitude, at least
for g-mode pulsations, and update the previously-observed effective temperature -- rotation period
relation.

\end{abstract}


\section{Introduction}
Kepler space telescope data have been fundamental in advances in
asteroseismology. The original mission (K1) concluded with four years of
nearly-uninterrupted single-instrument data. No comparable data set
has ever previously been obtained. Kepler data do not suffer from
daytime gaps nor differing instrumentation or observing sites trying
to use multiple Earth longitudes to obtain complete coverage. There are
no atmospheric effects and no Earth-orbiting complications (such as
the South Atlantic Anomaly). The follow-on mission, K2, has similarly
obtained heretofore unobtainable data sets, though of shorter (typically
about 80 days) duration. The K2 mission observed 20 fields %(with some overlap) 
in the ecliptic plane, ingeniously
using solar pressure to aid with pointing stability \citep{howell14a}.

Those data have resulted in nearly 1,600 publications on astrophysics
unrelated to Kepler's primary mission of discovering planets.
An important area of contribution is asteroseismology. Asteroseismology
uses pulsations to discern stellar structure and evolution. Prior to
Kepler, horizontal branch asteroseismology, where we can explore
compact, evolved cores undergoing helium fusion, was more effort than
result. There were debates whether oscillations had been detected at all in
red clump stars (solar-like oscillations).% and on the hot end, the
%extreme horizontal branch, where subdwarf B (sdB) and O (sdO) stars are
%($22,000\,\leq\,T_{\rm eff}\, \leq 40,000$K, $5.0\,\leq\,\log g\,\leq
%6.0$dex), 
Amongst hot  extreme horizontal branch stars, observations had done little to identify modes and hence 
constrain models.

Significant effort was expended trying to identify pulsation
modes in pulsating sdB (sdBV) stars using follow-up longer-duration 
photometry \citep[e.g.][]{reed07b}, multicolour photometry \citep[e.g.][]{randall06}, 
and time-resolved spectroscopy \citep[e.g.][]{baran10b}, with limited success. In large part, this was due to
ground-based observations' low-duty-cycle (typically $<30$\% coverage)
or limited duration on larger facilities. Kepler and K2's unique
data sets allowed observers to finally fully exploit these stars' pulsations
with seismic analyses.

Subdwarf B stars pulsate in both pressure (p) and gravity (g) modes; with the
hotter stars primarily p-mode dominated and the 
cooler stars predominantly
g-mode-dominated pulsators. Typical p modes have periods of a few
minutes with amplitudes rarely above 10 parts-per-thousand (ppt) while
g modes have longer periods, typically 1-4\,hours, with slightly lower
amplitudes. The variable star classifications for sdBV stars are
V361\,Hya for p mode pulsators \citep{kilkenny97}, 
V1093\,Her for g mode pulsators \citep{green03},  and DW\,Lyn \citep{schuh06} 
for hybrid pulsators which pulsate in both p and g modes. These classifications
have become less distinct as the majority of Kepler-observed sdBV stars have both types of pulsations.
DW\,Lyn as a hybrid has two strong p-mode pulsations and only one g-mode pulsation
yet for most of the Kepler-observed sdBV stars, the situation is reversed with more g modes observed
than p modes. To distinguish which type of pulsation is dominant (in amplitude and number of
detected pulsation perodicities), in this paper we use p+g and
g+p for p and g-mode dominated hybrid pulsators, respectively.
 For a review of sdB and sdO stars, see \citet{heber16},
which includes some early Kepler results.

Seismic discoveries using K1 and K2 data obtained for sdBV stars include
mode identifications using asymptotic g-mode-overtone period spacings
\citep{reed11c} and rotationally-induced frequency multiplets \citep{baran12a}.
These two methods have provided around two thousand identified modes;
to date, \emph{every} Kepler-observed sdBV star which has been analyzed
has had the majority of its periodicities associated with pulsation
modes (where $n$ represents radial overtones, $\ell$ the number of
surface nodes, and $m$ the azimuthal surface nodes). The most recent review of
Kepler results for sdBV stars is \citet{mdr_sdob8} and we examine group
properties in \S 4.

In this paper we analyze four sdBV stars discovered during K2's Campaign 7 (C7)
and place them in context with what has been detected so far.
\eone\, (2MASS\,19413850-2333426, GALEX J194138.5-233342) has $K_M=16.3$, % J19416-2333
\etwo\, (GALEX J191534.6-205107) has $K_M=16.3$, % J19155-2051
\ethree\, (2MASS\,19345376-1855522, GALEX J193453.7-185552) has $K_M=15.9$, and % J19348-1855
\efour\, (2MASS\,19334689-1817137, GALEX J193346.9-181713) has $K_M=15.8$ % 19337-1817

These four stars were part of our K2 Guest Observer program which observed
nine stars during C7, with these four pulsating. None of these stars were previously
known to pulsate, making them new discoveries.

In \S 4 we will discuss how these stars fit with what has been
learned from Kepler and published TESS \citep[Transiting Exoplanet Survey Satellite][]{tess1} 
data to date with a brief review. As this is our
47$^{th}$ paper using Kepler data, and with recent TESS
publications by \citet{charp19,shoaf20}, and \citet{sahoo20},
it is a good time to take an updated examination of group properties.

\section{Spectroscopic observations and results}
As part of our follow-up spectroscopic survey \citep{jhtsdob6},
low-resolution spectra (R $\sim$ 2000) have been collected for \eone,
\etwo, \ethree, and \efour\,
using the 2.56m Nordic Optical Telescope (NOT) with ALFOSC, grism \#18
and a 0.5 arcsec slit. We used CCD\#14 giving an approximate wavelength
range of 3450 -- 5350\AA , and a resolution based on the width of arc lines
of 2.2\AA.  Exposure times of 900\,s were used for all spectra.

The spectra were reduced and analyzed in the same way.  Standard
reduction steps within IRAF include bias subtraction, removal of
pixel-to-pixel sensitivity variations, optimal spectral extraction,
and wavelength calibration based on helium arc-lamp spectra. The
target spectra and the mid-exposure times were shifted to the
barycentric frame of the solar system.  Radial velocities (RVs) were derived
with the FXCOR package in IRAF. The RVs were
adjusted for the position of the target in the slit, judged from slit
images taken just before and after the spectral exposure.

The spectra of all four targets have the characteristic
appearance of single sdB stars, for which we cannot exclude binarity
with companions of much lower luminosity, such as main-sequence M
stars or white dwarfs.  For \etwo\, no 2MASS magnitudes are
listed.  For the other 3 targets there is no clear near infrared excess observed
(from 2MASS), again indicating single stars.  Nevertheless, for
\ethree\, we detect clear RV variability (see below).

\subsection{\eone, \etwo, and \efour}

For \eone, \etwo, and \efour, we determined the RV
 with the average spectrum of each target as a template
spectrum.  We also used those average spectra to derive the
atmospheric parameters of the stars, and we list these parameters in
Table\,1. %\ref{stable1}.  
For this purpose we used the fitting procedure of \citet{edel03}
with the metal-line blanketed local thermodynamic
equilibrium (LTE) models of solar composition described in \citet{heber00}.

For \eone\, we obtained nine useful spectra between Oct 2016 and Aug
2017, and achieved a signal-to-noise (S/N) level between 24 and 60
with median S/N=34.  The average spectrum has S/N$\approx$100.
The median RV error is $8.4\,{\rm km^{\cdot} s^{-1}}$  and the RV root-mean-square (RMS) of the individual
spectra around the average velocity is $7.1\,{\rm km^{\cdot} s^{-1}}$.
%## Only J-band 2MASS in EPIC, no SIMBAD info ##
%## J-V=0.2

For \etwo\, we obtained 10 useful spectra between Jun 2016 and Aug
2017, and achieved a S/N level between 29 and 57
depending on observing conditions, with median S/N=46.  The average
spectrum has S/N$\approx$110.   
The median RV error is $9.9\,{\rm km^{\cdot} s^{-1}}$  and the RV RMS of the individual
spectra around the average velocity is $10.5\,{\rm km^{\cdot} s^{-1}}$.
%## No 2MASS mags in EPIC, no SIMBAD info ##

For \efour\, we obtained eight useful spectra between Oct 2016 and Aug
2017, and achieved a S/N level between 44 and 74,
with median S/N=68.  The average spectrum has S/N$\approx$150.
The median RV error is $5.3\,{\rm km^{\cdot} s^{-1}}$  and the RV RMS of the individual
spectra around the average velocity is $7.8\,{\rm km^{\cdot} s^{-1}}$.
%## Only JH-bands 2MASS in EPIC, no SIMBAD info ##
%## J-V=0.4, H-V=0.3

We conclude that for these three targets our RV measurements are
consistent with single stars.

\begin{table}
\label{stable1}
\caption{Results of spectral analysis. Errors on the final digits are given in parentheses.}
\begin{tabular}{lccc} \hline
Star &     $T_{\rm eff}$ (K) &       $\log g$ (cgs) &   $\log \left(\frac{N_{He}}{N_H}\right)$\\
\eone &  27860(160) & 5.45(2) &  -2.718(38)\\
\etwo &  22770(150) & 5.01(2) &  -2.104(77)\\
\ethree &  28160(110)&  5.44(2) &  -2.862(28)\\  %% Orbit corrected average
\efour &  24470(160) & 5.17(2) &  -2.633(61)\\
\hline
\end{tabular}
\end{table}

\subsection{ The orbit of \ethree}

For \ethree\, we obtained 12 useful spectra between Oct 2016 and Oct
2017, and achieved a S/N level between 34 and 71,
with a median S/N=51.  The average spectrum has S/N$\approx$140.
For determining the RVs we have first used the average spectrum as a
  cross-correlation template, and subsequently a spectral model
  (as in Table\,2) as a template. %\ref{spectable2}) as a template.
The median RV error is $7.7\,{\rm km^{\cdot}s^{-1}}$.

We find significant RV variations, and list the orbital solution
obtained while assuming a circular orbit, in Table\,2.  With an %\ref{spectable2}.  With an
orbital-velocity amplitude of $66.3\, {\rm km^{\cdot}s^{-1}}$, an orbital period of 5.92
days, and assuming that the companion is an unseen white dwarf, we
derive the following constraints from the mass function. For a
canonical mass of the subdwarf of 0.47 M$_{\odot}$ \citep{vvg2014} the
minimum mass of the WD companion is 0.58 M$_{\odot}$.  For an ensemble of
a canonical-mass sdB with a canonical-mass WD (0.6 M$_{\odot}$) the orbital
inclination is 79$^o$.  Assuming a mass of 0.3 M$_{\odot}$ for the sdB
the minimum mass of the WD is still 0.48 M$_{\odot}$, and the minimum orbital
separation is 12.6 R$_{\odot}$. Using canonical masses and $i=90^o$, the separation
is 14.3R$_{\odot}$. At that separation and using R$_{\rm sdB}=0.2$R$_{\odot}$,
eclipses would occur for $i>89^0$. No eclipses are observed in the K2 data (\S 3.3)
indicating $i<89^o$.

%%%%%%%%%%%%%%%%%%%%%%%%%%%%%%%%%%%%%%%%%%%%%%
\begin{table}
\label{spectable2}
\caption{Results of spectral binary analysis for \ethree.}
\begin{tabular}{lc} \hline

\multicolumn{2}{c}{Solution of RV wrt mean spectrum} \\
 Amplitude &                $64.01   \pm  3.09\,{\rm km^{\cdot}s^{-1}}$ \\
 Period    &                 $5.9190 \pm  0.0021\,$d \\
 Reduced $\chi^2$  &             1.408 \\
 RMS               &         $8.8\,{\rm km^{\cdot}s^{-1}}$  \\ \hline

\multicolumn{2}{c}{Solution of RV wrt model template} \\
 System velocity   &        $38.12   \pm  3.51 \,{\rm km^{\cdot}s^{-1}}$\\
 Amplitude         &        $66.28   \pm  3.43\,{\rm km^{\cdot}s^{-1}}$ \\
 Period            &         $5.9218 \pm  0.0026$\,d \\
 Reduced $\chi^2$  &             0.73 \\
 RMS               &         $8.8\,{\rm km^{\cdot}s^{-1}}$ \\ \hline

\multicolumn{2}{c}{Solution of RV wrt model template, with all RV errors set to $9.7\,{\rm km^{\cdot}s^{-1}}$}\\
 System velocity   &        $38.20   \pm  3.15 \,{\rm km^{\cdot}s^{-1}}$\\
 Amplitude         &        $70.95   \pm  4.03 \,{\rm km^{\cdot}s^{-1}}$\\
 Period            &        $ 5.9187 \pm  0.0024$\,d \\
 Reduced $\chi^2$  &             1.00 \\
 RMS               &         $7.9\,{\rm km^{\cdot}s^{-1}}$ \\
\hline
\end{tabular}
\end{table}
%%%%%%%%%%%%%%%%%%%%%%%%%%%%%%%%%%%%%%%%%%%%%%

To determine the atmospheric parameters of \ethree, we shifted all
individual spectra to remove the orbital velocities, and made an
average orbit-corrected spectrum, to which we applied the same
modelling procedure as described above for the other stars. The fit
results are presented in Table\,1. %\ref{stable1}.

\begin{figure}
\centerline{\psfig{figure=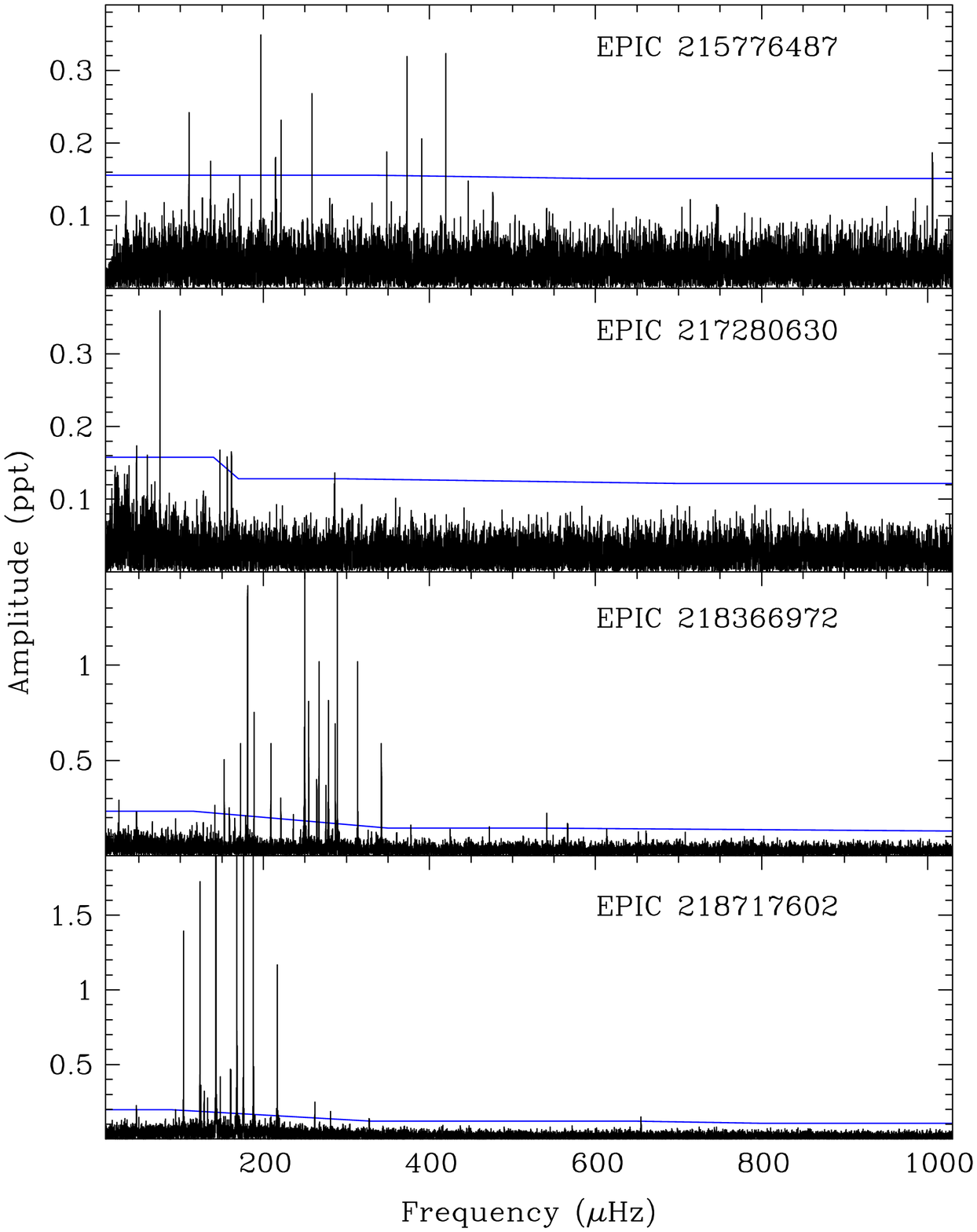,width=\columnwidth} }
\caption{Fourier transforms of \eone , \etwo , \ethree , and \efour .
Horizontal (blue on-line) lines indicate the detection threshold.}
\label{figFT4b}
\end{figure}

\begin{table*}
\label{tabeone}
\caption{Period list for \eone. $^{\dagger}$: This frequency was fitted using
data from days 25-40 of the run only with $\sigma\,=\,0.086$\,ppt.} %E2157
% done Dec. 27, 2018
\begin{tabular}{lccccccc} \hline
ID & Freq & Per & Amp & S/N & $\ell$ & $n$ & $\frac{\Delta P}{P}$ \\
   & $\mu$Hz & Sec & ppt &     &     & & \% \\
f1 & 110.384 (8)    & 9059.31 (67) & 0.27  (3) & 7.3 & 1 & 33    & 3.2 \\
f2 & 136.606 (10)    & 7320.30 (51) & 0.23  (3) & 6.3 & 1    & 26 &  0.03 \\  
f3 & 171.452 (13) & 5832.53 (46)   & 0.16 (3)  & 4.4 &  1    & 20 & -1.6 \\
f4 & 197.027 (6)    & 5075.45 (16) & 0.35  (3) & 9.8 &1/2 & 17/32 & -7.7/-5.3 \\
f5 & 214.565 (10)    & 4660.59 (22) & 0.22  (3) & 6.2 &2    & 29    & 4.2 \\
f6 & 221.276 (10)    & 4519.24 (21) & 0.22  (3) & 6.2 & 2    & 28    & 5.1 \\
f7 & 258.647 (8)    & 3866.28 (12) & 0.27  (3) & 7.6 &1    & 12 &  3.3 \\  
f8 & 348.718 (12)    & 2867.65 (10) & 0.19  (3) & 5.3 & 1    & 8 &   -0.5 \\  
f9 & 372.872 (7)    & 2681.88 (5) & 0.32  (3) & 9.0&  &  &     \\     
f10 & 390.700 (11)    & 2559.51 (7) & 0.20  (3) & 5.7 &1    &   &    \\
f11 & 419.783 (7)    & 2382.18 (4) & 0.32  (3) & 9.0 &1/2    & 6/13 &   3.2/8.4 \\ 
f12$^{\dagger}$ & 1005.921 (71)   & 994.11 (7) & 0.41  (7)     & 4.8 \\
\hline
\end{tabular}
\end{table*}

\begin{table*}
\label{tabetwo}
\caption{Period list for \etwo. $^{\dagger}$ Values for f6 are from Lorentzian fitting.} %E2172
% done Dec. 28, 2018
\begin{tabular}{lccccccc} \hline
ID & Freq & Per & Amp & S/N & $\ell$ & $n$ & $\frac{\Delta P}{P}$ \\
   & $\mu$Hz & Sec & ppt &     &     & & \% \\
f1 & 60.259 (8) & 16549.39 (2.18) & 0.144 (23) & 6.2 & 1 & 80 & -3.7 \\
sA & 65.295 (12) & 15315.08 (2.81) & 0.140 (22) & 3.9 & 1 & 70 & 1.1 \\
f2 & 75.454 (62) & 13253.04 (10.84) & 0.390 (22) & 11.0 & 1 & 63 & 6.6 \\
f3 & 147.652 (65) & 6772.68 (2.96) & 0.184 (22) & 7.1 & -- & -- & -- \\
f4 & 156.544 (47) & 6387.98 (1.91) & 0.178 (22) & 6.1 & 1 & 28 &-4.3 \\
f5 & 161.638 (54) & 6186.66 (2.06) & 0.153 (22) & 5.9 & 1 & 27 & -1.4 \\
f6$^{\dagger}$ & 285.85 (18) & 3498.31 (2.23) & 0.107 & -- & 1&13&  2.7  \\
\hline
\end{tabular}
\end{table*}

\section{K2 observations and pulsation analyses}
Campaign 7 spanned 82\,days between 4 October and 26 December, 2015.
Our data are short-cadence observations, with integration times of 58.85\,seconds
which were downloaded from MAST as pixel files. Fluxes were extracted using
aperture photometry and spacecraft artefacts were removed using our
custom process described in \citet{baran16d,ketzerF2}. Temporal spectra
(Fourier transforms, FTs; Fig.\,\ref{figFT4b}) were produced to examine
the pulsations and sliding FTs (SFTs) were produced to examine the
time-stability of the pulsations. The 1.5/T frequency resolution
of these data is
$2.14\,\mu$Hz and to make it unlikely any peak in the FT is due to random
noise requires a S/N of $4.2\sigma$, where $\sigma$
is the average level of the FT. As low frequency noise is more difficult to
remove, $\sigma$ were calculated in frequency regions nearly devoid of pulsations
and linearly interpolated between, where pulsations occur.

%\begin{figure*}
%\centerline{\psfig{figure=ft4F7.ps,angle=-90,width=\textwidth} }
%\caption{Fourier transforms of .
%Dashed vertical lines indicate known spacecraft artefacts and solid
%vertical line indicates the Nyquist frequency.}
%\label{figFT4}
%\end{figure*}

\begin{table*}
\label{tabethree}
\caption{Period list for \ethree. $^{\dagger}$ frequencies were Lorentzian fitted.
$^{\star}$ indicates mode identifications which are less certain.} %E2183
% done Dec. 28, 2018
\begin{tabular}{lcccccccc} \hline
ID & Freq & Per & Amp & S/N & $\ell$ & $n$ & $\frac{\Delta P}{P}$ \\
   & $\mu$Hz & Sec & ppt &     &     & & \% \\
sA & 119.945 (10) & 8337.14 (71) & 0.203 (26) & 4.0 & 1/2 & 29/53 & -0.5 / -6.0 \\
sB & 127.638 (11) & 7834.66 (66) & 0.191 (26) & 3.8 & 1 & 27 & 2.8 \\
sC & 136.773 (11) & 7311.39 (61) & 0.183 (26) & 3.7 &1/2 & 25/46 & -2.1 / -2.9 \\
f04 & 141.552 (07) & 7064.53 (37) & 0.297 (28) & 5.9 & 1 & 24 & 1.3 \\
f05 & 152.926 (4) & 6539.13 (17) & 0.548 (28) & 11.3 & 1 & 22 & -4.4 \\
f06 & 159.091 (8) & 6285.72 (32) & 0.267 (28) & 5.8 & 1/2 & 21/39 & -3.6 / 0.3 \\
sD & 165.698 (10) & 6035.08 (37) & 0.126 (28) & 4.0 & 1 & 20 & -1.7 \\
f08 & 172.580 (4) & 5794.43 (12) & 0.591 (28) & 12.5 & 1 & 19 & 4.1 \\
f09 & 178.679 (10) & 5596.61 (30) & 0.212 (28) & 4.5 &  &  &  \\
f10 & 181.005 (2) & 5524.70 (5) & 1.398 (28) & 29.6 &1 & 18 & -1.5 \\
f11 & 188.826 (3) & 5295.88 (8) & 0.766 (26) & 16.2 & &  &  \\
f12 & 209.051 (3) & 4783.53 (8) & 0.597 (26) & 13.2 & 1 & 15 & 8.4 \\
f13 & 221.076 (7) & 4523.34 (14) & 0.309 (26) & 6.8 & 1/2 & 14/27 & 6.5 / 3.0 \\
f14 & 236.030 (8) & 4236.75 (15) & 0.252 (26) & 5.8 & 1/2 & 13/25 & -5.7 / 8.3 \\
f15$^{\dagger}$ & 244.74 (23) & 4086.1 (3.8) & 0.16 (--) & -- & 2 & 24 & 5.0 \\
f16 & 249.782 (1) & 4003.48 (1) & 2.742 (26) & 62.3 & 1 & 12 & 3.0 \\
f17 & 254.659 (2) & 3926.82 (4) & 0.832 (26) & 18.9 & 2 & 23 & -2.2 \\
f18 & 264.374 (5) & 3782.51 (7) & 0.439 (26) & 18.9 & 2 & 22 & -0.2 \\
f19 & 267.029 (2) & 3744.92 (3) & 1.020 (28) & 10.0 & 1 & 11 & 1.8 \\
f20 & 275.184 (6) & 3633.94 (8) & 0.370 (28) & 23.2 & 2 & 21 & -1.2 \\
f21 & 278.242 (4) & 3593.99 (5) & 0.637 (28) & 14.4 &   &  &  \\
f22 & 278.543 (3) & 3590.10 (4) & 0.769 (28) & 17.5 & &  &  \\
f23$^{\dagger}$ & 283.21 (26) & 3531.0 (3.2) & 0.13 (--) & -- &  &  \\
f24 & 286.534 (3) & 3489.98 (4) & 0.743 (28) & 18.1 & 1/2 & 10/20 & -2.0 / 1.0 \\
f26 & 289.181 (1) & 3458.04 (2) & 1.538 (28)  & 37.5 & 1 & 10 & -10.5 \\
f27 & 313.565 (2) & 3189.14 (2) & 1.023 (28) & 25.8 & 1/2 & 9/18 & -0.3 / -3.4 \\
sE & 326.205 (12) & 3065.56 (12) & 0.168 (26) & 4.3 & $^{\star}$2 & 17 & 12.7 \\
f29 & 330.446 (12) & 3026.21 (11) & 0.177 (26) & 4.6 & $^{\star}$2 & 17 & -14.1 \\
f30 & 341.948 (4) & 2924.42 (3) & 0.452 (28) & 11.6 &1 & 8 & 1.1 \\
f31 & 342.134 (5) & 2922.83 (4) & 0.489 (28) & 12.5 & 1 & 8 & 0.5 \\
sF & 360.526 (16) & 2773.73 (12) & 0.129 (26) & 3.9 &  &  &  \\
f33 & 377.650 (10) & 2647.95 (7) & 0.188 (25) & 6.0 & 1 & 7 & -1.9 \\
f34 & 424.882 (13) & 2353.59 (7) & 0.151 (25) & 4.5 &   &  &  \\
f35 & 472.046 (12) & 2118.44 (5) & 0.165 (25) & 5.2 & 1 & 5 & 0.9 \\
f36 & 541.427 (9) & 1846.97 (3) & 0.241 (26) & 7.2 & 1 & 4 & -0.3 \\
sG & 613.733 (13) & 1629.37 (4) & 0.146 (25) & 3.9 &  &  &  \\
sH & 651.577 (12) & 1534.74 (3) & 0.164 (25) & 3.9 & &  &  \\
sI & 660.887 (16) & 1513.12 (4) & 0.137 (25) & 4.3 & &  &  \\
sJ & 661.036 (17) & 1512.78 (4) & 0.111 (25) & 4.0 &  &  &  \\
sK & 708.144 (12) & 1412.14 (2) & 0.158 (25) & 4.1 &  &  \\
\hline
\end{tabular}
\end{table*}

\begin{table*}
\label{tabefour}
\caption{Period list for \efour. $^{\dagger}$ f18 was nlls fitted
using only the first 10 days of data with $\sigma\,=\,0.07$\,ppt.} %E2187
% done Dec. 29, 2018
\begin{tabular}{lccccccc} \hline
ID & Freq & Per & Amp & S/N & $\ell$ & $n$ & $\frac{\Delta P}{P}$ \\
   & $\mu$Hz & Sec & ppt &     &     & & \% \\
f01 & 103.904 (2) & 9624.29 (14) & 1.376 (27) & 40.5 & 1 & 33 & 0.1 \\
f02 & 123.749 (1) & 8080.89 (8) & 1.689 (27) & 49.7 &  &  &  \\
f03 & 124.653 (6) & 8022.30 (40) & 0.293 (27) & 8.6 & 1 & 27 & -8.7 \\
f04 & 128.692 (6) & 7770.48 (39) & 0.371 (27) & 11.0 &1 & 26 & -4.4 \\
f05 & 132.902 (8) & 7524.35 (41) & 0.264 (27) & 7.8 & 1/2 & 25/46 & 2.1/-4.7 \\
f06 & 142.845 (1) & 7000.58 (4) & 2.400 (28) & 70.6 & 1 & 23 & 3.0 \\
f07 & 143.045 (1) & 6990.82 (6) & 1.760 (28) & 51.8 & 1 & 23 & -0.7 \\
f08 & 148.202 (5) & 6747.56 (22) & 0.431 (27) & 12.7 &1 & 22 & 6.9 \\
f09 & 160.668 (5) & 6224.03 (18) & 0.451 (27) & 13.4 & 1 & 20 & 7.9 \\
f10 & 168.133 (1) & 5947.67 (2) & 3.710 (27) & 112.4 & 1 & 19 & 2.9 \\
f11 & 176.074 (1) & 5679.44 (3) & 1.989 (27) & 60.3 &  1 & 18 & 1.0 \\
f12 & 187.938 (1) & 5320.91 (3) & 2.127 (27) &  64.5 &  &  &  \\
f13 & 216.857 (2) & 4611.33 (4) & 1.121 (27) & 35.0 & 1 & 14 & -4.9 \\
f14 & 261.915 (9) & 3818.03 (14) & 0.227 (27) & 7.1 &  1 & 11 & -6.4 \\
f15 & 280.983 (9) & 3558.93 (15) & 0.169 (27) & 5.4 &  1 & 10 & -4.9 \\
f16 & 327.077 (11) & 3057.39 (10) & 0.129 (18) & 4.8 & 1 & 8 & 4.5 \\
f17 & 327.867 (10)& 3050.02 (10) & 0.135 (18) & 5.0 & 1 & 8 & 1.7 \\
f18$^{\dagger}$ & 7876.05 (14) & 126.967 (2) & 0.368 & 5.3 & \\
f19 & 8110.90 (7)&   123.290 (1)     & 0.122 (43) & 4.5 & \\
\hline
\end{tabular}
\end{table*}

Three of the four stars are exclusively g mode pulsators with only \efour\, having
two p-mode periodicities. Most of the 
pulsations were amplitude and phase
stable, in which case we prewhitened them using non-linear least-squares (nlls) 
fitting \citep[e.g.][]{reed04b}. Otherwise, we either Lorentzian fitted the FTs, 
using the Lorentzian widths as an
estimator of frequency uncertainty only \citep[e.g.][]{reed14} or prewhitened smaller
sections of data when the pulsations were significantly above the noise. Pulsations we detected are
supplied in Tables\,3 through 6. When the amplitudes were
%\ref{tabeone} through \ref{tabefour}. When the amplitudes were
sufficiently stable for prewhitening, we include the fitting error in 
the tables, otherwise we do not.

While C7 spanned 82\,days, to detect rotationally-split frequency multiplets usually requires
two rotations within the observations. For stars where we do not detect frequency
multiplets, we presume a spin period $>45$\,days. Of the 18 K1-observed sdBV stars,
which had multiple years of data, eight (44\%) had spin periods $\geq 45$\,days.

\subsection{Pulsation analyses of \eone}
A total of 12 periodicities were detected above the detection threshold (shown as
a horizontal blue line in Fig.\,\ref{figFT4b}) between 110 and $1006\,\mu$Hz
(994 and 9058\,s). These were all stable in amplitude/frequency and so were
nlls fitted. Their frequencies, periods, amplitudes, and S/N are provided in
Tab.\,3. %\ref{tabeone}.

\begin{figure*}
\centerline{\psfig{figure=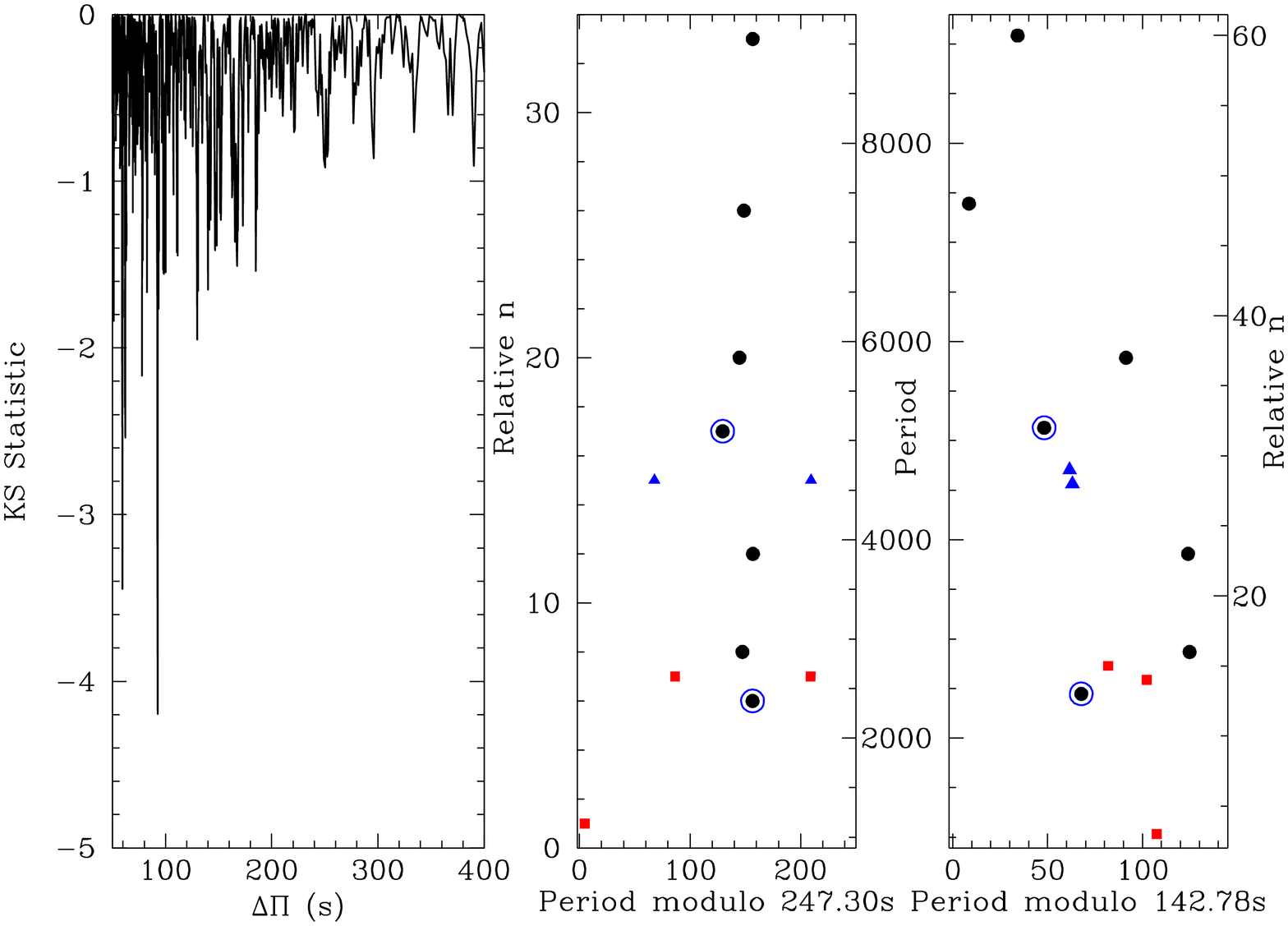,angle=-90,width=\textwidth} }
\caption{KS test (left panel) and echelle diagrams (right two panels) for
$\ell\,=\,1$ and 2 asymptotic sequences of \eone , respectively. In the echelle
diagrams, black circles indicate periods that match the $\ell\,=\,1$ sequence, blue triangles
match the $\ell\,=\,2$ sequence, black circles with a blue surround match both sequences, and red points
do not fit either sequence.}
\label{ks1}
\end{figure*}

For g modes in sdB stars, 
typical $\ell =1$ asymptotic period sequences have been found to have spacings ($\Pi_{\ell =1}$)
near 250\,s, and even a cursory differencing of \eone 's periods reveals
similar spacings. In usual fashion, we do a Kolomogorov-Smirnof (KS) test,
which can reveal commonly spaced periods. Very surprisingly for \eone\,
there is no signature trough near 250\,s to indicate the sequence
(left panel of Fig.\,\ref{ks1}). Another tool
used to discover asymptotic period sequences is an echelle diagram. We produced
one with a spacing of 250\,s and the sequence appeared
(middle panel of Fig.\,\ref{ks1}). Using that as our guide,
we found eight periods that are part of the $\ell =1$ sequence and these modes 
indicate a
period spacing of $247.3\,\pm\,0.4\,s$. From the $\ell =1$ sequence, an
$\ell =2$ sequence can be calculated from the relation $\Delta \Pi_{\ell =2}
=\Delta \Pi_{\ell =1}/\sqrt{3}$. Two periods were found to fit
the $\ell =2$ sequence and two of the $\ell =1$ periods could also fit the
$\ell =2$ sequence, and are marked accordingly in Tab.\,3. %\ref{tabeone}.
All but two of the frequencies fit these sequences.
f9 does not fit either sequence, though it is one of the highest-amplitude
periodicities. Perhaps this is a trapped mode, though we cannot confirm
this as our series are not contiguous in $n$, which would be a necessary condition
for finding trapped modes. 
There are no evenly-split frequencies indicative of rotationally-split
multiplets and so we cannot determine a rotation period for \eone . It
is likely longer than our sensitivity, which is about 45\,d.

\begin{figure}
\centerline{\psfig{figure=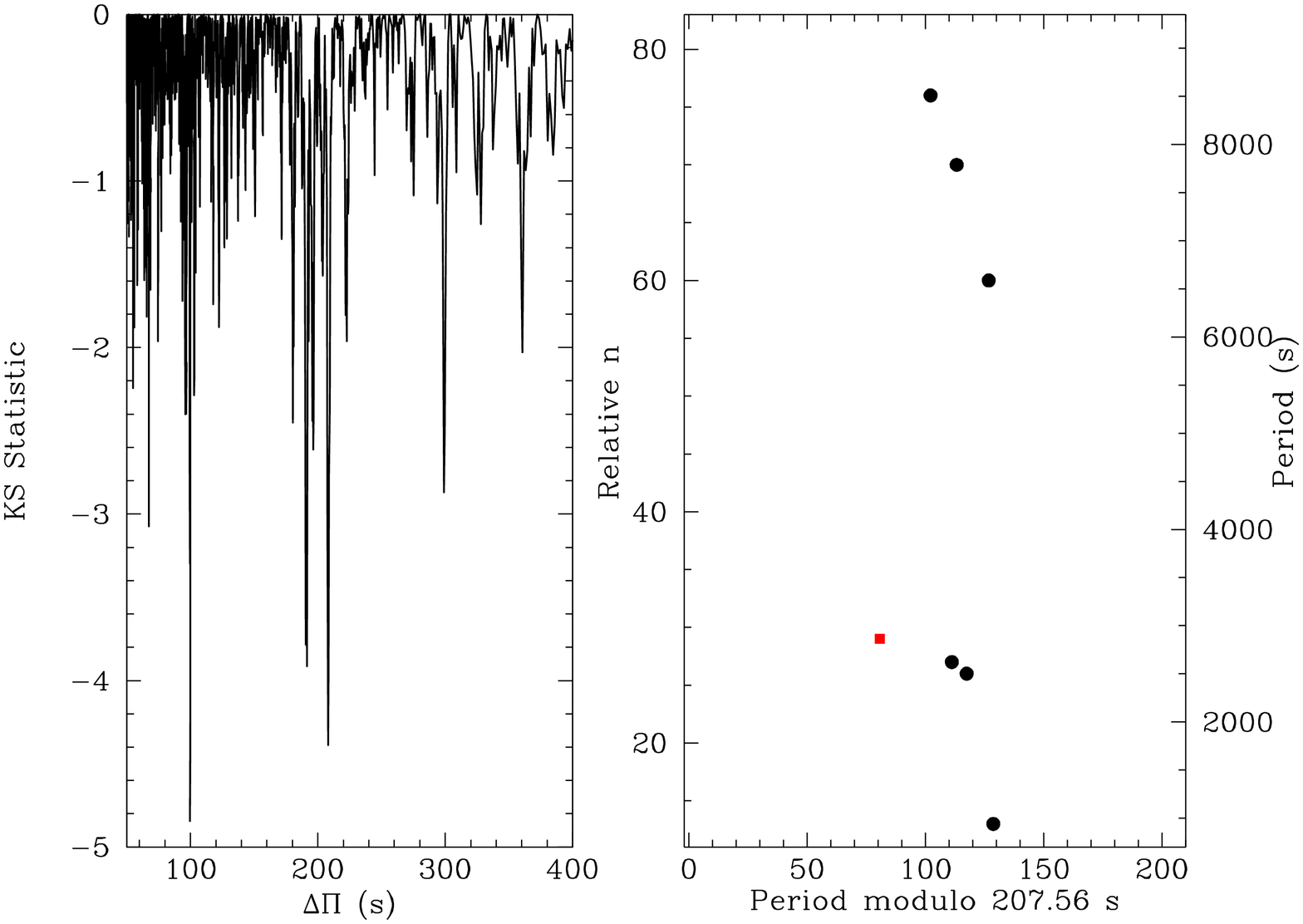,angle=-90,width=\columnwidth} }
\caption{KS test (left panel) and $\ell =1$ echelle diagram (right panel) for \etwo .
Point shapes and colours (on-line only) as in Fig.\,\ref{ks1}.}
\label{ks2}
\end{figure}

\subsection{Pulsation analyses of \etwo}
We only detect six frequencies in \etwo 's data set and two of these are very
low near 60 and 75\,$\mu$Hz, which must be very close to the acoustic cut-off. All
except for f6 were nlls fitted.
A KS test has a significant trough just past 200\,s and the echelle diagram confirms
that sequence (both shown in Fig.\,\ref{ks2}). 
An additional low-amplitude peak (labeled sA in Tab.\,4, %\ref{tabetwo}),
which is obvious in the FT, below the detection threshold, but fits the asymptotic
sequence is included in both the figure and the table for \etwo . Linear regression
determines $\Pi_{\ell =1}=207.56\,\pm\,0.26\,s$
  for \etwo , with only f3 not fitting
the $\ell =1$ sequence. Like \eone , this non-sequence periodicity has a fairly high
amplitude (the second highest) and so could represent a trapped $\ell =1$ mode,
though we have no way to confirm this. The sequence has only one contiguous pair,
and so there is no way to search for trapped modes. The period spacing of
207\,s is extreme for a cool (coolest of these four stars), 
purely-g-mode pulsator. There are two other sdBV
stars with small $\Pi_{\ell =1}$ values, but they are much hotter, p+g hybrid
pulsators. There are no indicators
of rotationally-induced frequency multiplets which we interpret as a long ($>45\,d)$
rotation period. However, it could also indicate an orientation where the
$m\neq 0$ components are of lower amplitude, and therefore undetected.

\begin{figure*}
\centerline{\psfig{figure=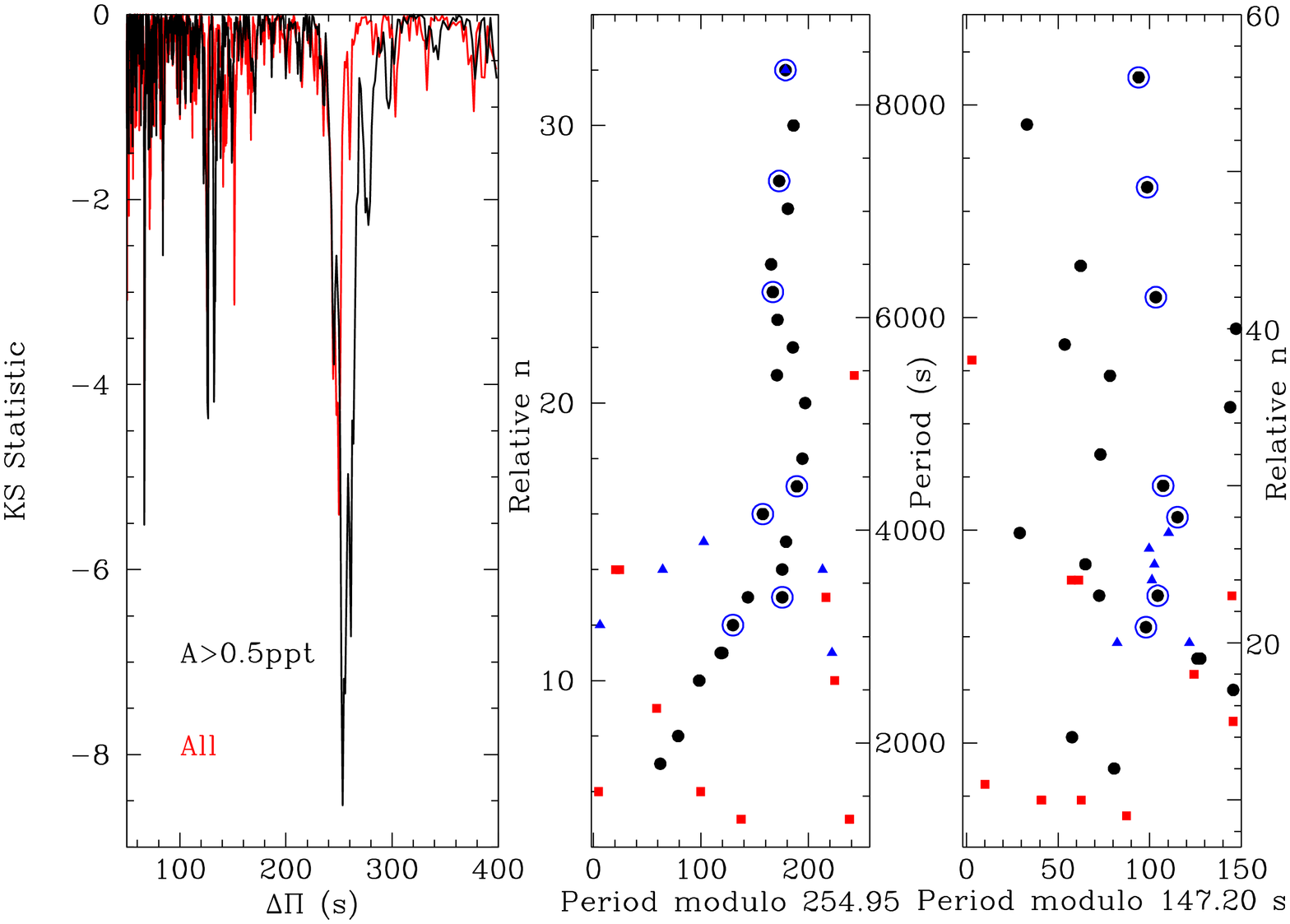,angle=-90,width=\textwidth} }
\caption{Same as Fig.\,\ref{ks2} for \ethree .}
\label{ks3}
\end{figure*}

\subsection{Pulsation analyses of \ethree}
\ethree\, is the richest pulsator of the group. We detect 36 periodicities
above the detection limit with another 11 \emph{suspected}. The strongest trough
in the KS test (left panel of Fig.\,\ref{ks3})
is near 250\,s, though it is not especially significant compared
to others. As $\ell =1$ modes have the least geometric cancellation \citep{pes85} and therefore
typically higher amplitudes, we did an amplitude cut at 0.5\,ppt and, with just those periods,
the trough became significant.% As $\ell =1$ modes have the least geometric cancellation
%\citep{REF}, removing low-amplitude periods should favor the $\ell =1$ sequence.
The echelle diagram (right panel of Fig.\,\ref{ks3}), 
spaced at 254\,s, easily shows that sequence, however \ethree\,
has a significant ``hook'' feature at lower radial overtones. This has been seen
in several other sdBV stars \citep{baran12c}. We calculated a
period spacing of $254.95\,\pm\,0.50\,s$ with linear regression above the hook,
and then linearly fitted the hook feature for inclusion in Tab.\,5. %\ref{tabethree}.
We calculated where $\ell =2$ sequence periodicities should occur from the
$\ell =1$ sequence and those that match are labelled as such in Tab.\,5. %\ref{tabethree}.
Many of the $\ell =1$ modes also match the $\ell =2$ sequence, and since
we cannot distinguish between them, we label them with both. An intriguing feature is how linear
the periods look below the turn of the ``hook'' feature ($<4\,000$\,s). A KS test of just
those periods reveals a broad peak with a central value of 271\,s and folding across that
period produces an echelle with seven periods in line. As such, an alternative view would be
that there is a sequence with periods $>3\,450$\,s with an asymptotic spacing of 254.95\,s and
another at 271\,s up to and including f26 (3\,458\,s).

Surprisingly for so many
frequencies there are no obvious rotationally-induced frequency
multiplets. If \ethree\, were tidally locked, a rotation period of 5.9\,d
would be a frequency of about $2.0\mu$Hz and $\ell =1$ modes would have
a separation of half that near $1.0\mu$Hz. There are no frequency splittings
near those values or multiples. The closest we find to a multiplet would be
a possible triplet of f15-f16-f17, which are split by 4.95 and 4.88$\mu$Hz,
respectively. However, those three periods also fit asymptotic sequences, making
that a more likely fit. As such, we do not detect any frequency multiplets
in \ethree . In this case, binarity indicates an orientation favorable for
viewing multiplets, and so we can safely assume that the rotation period is $>45$\,d.

\begin{figure}
\centerline{\psfig{figure=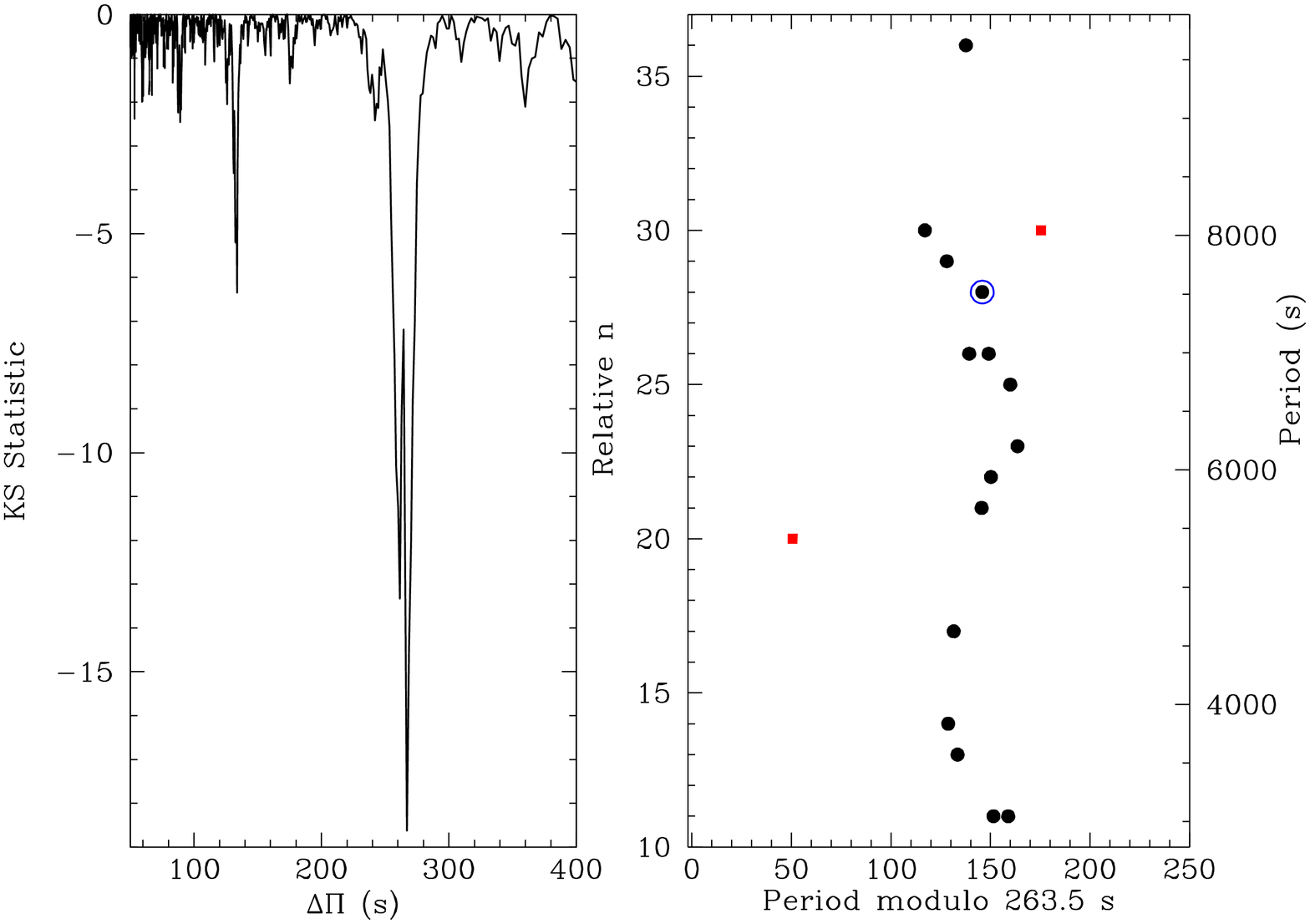,angle=-90,width=\columnwidth} }
\caption{Same as Fig.\,\ref{ks2} for \efour .}
\label{ks4}
\end{figure}

We can also use the FT to search for a signal from Doppler boosting caused
by the binary motion. Following the method of \citet{telting12a} and using
K$\,=\,67$\,km$\cdot$s$^{-1}$, we calculate a Doppler boosting signal of
0.3\,ppt at $1.96\,\mu$Hz. The final step in our lightcurve processing
removes trends $>1.5$\,days, and so we examined a lightcurve without
this processing step. Unfortunately, the unprocessed lightcurve has low-frequency 
noise greater than 0.3\,ppt and so we cannot detect the binary signal using
Doppler boosting. There is also no indication of eclipses in the lightcurve.

\subsection{Pulsation Analyses of \efour}
We were able to prewhiten 19 frequencies for \efour\, and, as is obvious just by
looking at the FT, most of these readily fit into an $\ell =1$ asymptotic
sequence. The KS test shows a very deep trough near 265\,s, which is easily
reproduced in the echelle diagram (both are shown in Fig.\,\ref{ks4}). There is a slight
hook feature, but a linear regression finds a
period spacing of $263.15\,\pm\,0.48\,s$. This sequence includes all but four
of the g modes. Of those, there are two pairs (f02, f03 \& f16,f17) 
separated by $0.84\pm 0.08\mu$Hz. If $\ell =1$ (as f03, f16, and f17 fit the sequence)
then this would be a rotation period of about 7\,d. These are not the highest
amplitude pulsations in \efour\, and so the multiplet detection is not secure.
\efour\, also has two p-mode frequencies, with f18 only appearing at the beginning
of the run. However, without any observational
evidence for mode identifications, they do not tell us much. Only that another
quite cool sdB star has p modes (see Tab.\,7 for others). %\ref{tabgrp1} for others).

\section{Examining the group}
Kepler's original mission (K1) observed 18 sdBV stars \citep{roy10b,roy11b} and
K2 observed 139 of our proposed sdB targets in short-cadence mode during its 20 campaigns.
Many K1-observed sdBV stars have over three years, or about 1.5\,million data points,
of observations, while K2 only observed individual fields for roughly 80\,days, resulting
in about 110\,000 data points per target.
While this vast wealth of data takes quite some time to process (particularly
K2 stars for which we have to begin with pixel files), first-look analyses
anticipates about 50 pulsators from K2. To date a total of 34 of the roughly 69
Kepler-observed sdBV stars have
been analyzed and published (including the four in this paper), so there is
still some ways to go. Additionally, TESS has now completed its two year main mission, during
which it observed about 1\,000 of our proposed sdB targets, few of which have been examined as of
this writing.
As such, it seems a good point to examine progress
and compare and review what has been detected.

Thanks to a generous time allocation from the NOT, we have
been able to obtain spectra of all of our Kepler targets \citep{jhtsdob6}. We do this
both to constrain binarity and so that we are fitting the same resolution spectra
to the same atmospheric model grids. While there may be systematics between model
grids or differing instrumentation in data from multiple sources, our
single-sourced \emph{relative} values should be accurate.
We refer readers to any of the references provided in Tab.\,7 for details %\ref{tabgrp1} for details
on the spectra and their processing.
Table\,7 lists seismic, spectroscopic, and orbit-rotation 
properties of 38 published (or \emph{in press}) Kepler-observed sdBV stars,
 one blue horizontal branch (BHB) star with closely related pulsation properties,
and four TESS-observed sdBV stars.
Rotation is deduced strictly from pulsation frequency multiplets while the binary period
may be deduced from RV or photometric variations. To date, from K1 there were
two stars analyzed for which multiplets were not detected while K2 has many.
From K1, we know that rotation periods typically span tens of days, and so any periods
longer then about 45\,d (44\% of K1 stars) would not likely be resolved during K2 observations. Each TESS 
sector of observations spans about 26\,days, and so rotation periods longer than about 12\,days
(68\% of K1+K2 stars) would likely not be detected. In this ``group'' summary, we only consider
sdBV stars observed during K1, K2, and TESS missions and excluding atypical pulsation types
\citep[e.g.][]{csj17}. This sample should include stars with similar bulk physical properties that
have observations obtained in a roughly-homogeneous manner capable of providing mode identifications.
Two exceptions to this are the sdB+WD binaries Feige\,48 \citep{reed12c} and
KPD\,1930$+$2752 \citep{mdr11} included only in Fig.\,\ref{figbinspin} 
which have binary periods under one day and rotation periods derived from ground-based pulsation data.

The real revolution from Kepler data is the ability to observationally
correlate periodicities with pulsation modes (mode identifications).
The main tools which have been used are asymptotic g-mode period spacings,
which provide $\ell$ and relative $n$ values;
rotationally-induced frequency multiplets, which provide $\ell$, $m$, and
can provide relative $n$ if several multiplets of the same degree are detected;
and g-mode frequency multiplets splittings, which have relative spacings
dictated by the Ledoux constant \citep{led51}, and this can provide $\ell$
values.

These identifications
are invaluable for constraining stellar structure models, from which we
discern internal physics. Important pulsation properties include
the smoothness of asymptotic sequences \citep[e.g.][]{reed14},
which describe less stratified transition regions than
expected \citep{constantino15};
mode trapping \citep[e.g.][]{roy14a}, which has now been associated
with convective core overshoot \citep{guo2019,ost21}; both
p and g mode overtone spacings \citep{reed11c,baran12a},
which describe the resonant
cavity; and frequency multiplets, which provide information on rotation,
including differential rotation, and
for stars in binaries, which all indicate subsyncrhonous rotation
for close binaries \citep[periods under 10 days, e.g.][]{telting12a},
with the exception of the 3\,hour binary 2M1938+4603 \citep{2m1938},
constraining synchronization time scales for post-common-envelope (PCE)
binaries \citep[e.g.][]{pablo12}. Hybrid pulsators allow for radial scrutiny
as g modes probe deeper than p modes \citep{charp14}, with the
latter being mostly envelope (defined as above the He/H transition).

\begin{figure*}
\centerline{\psfig{figure=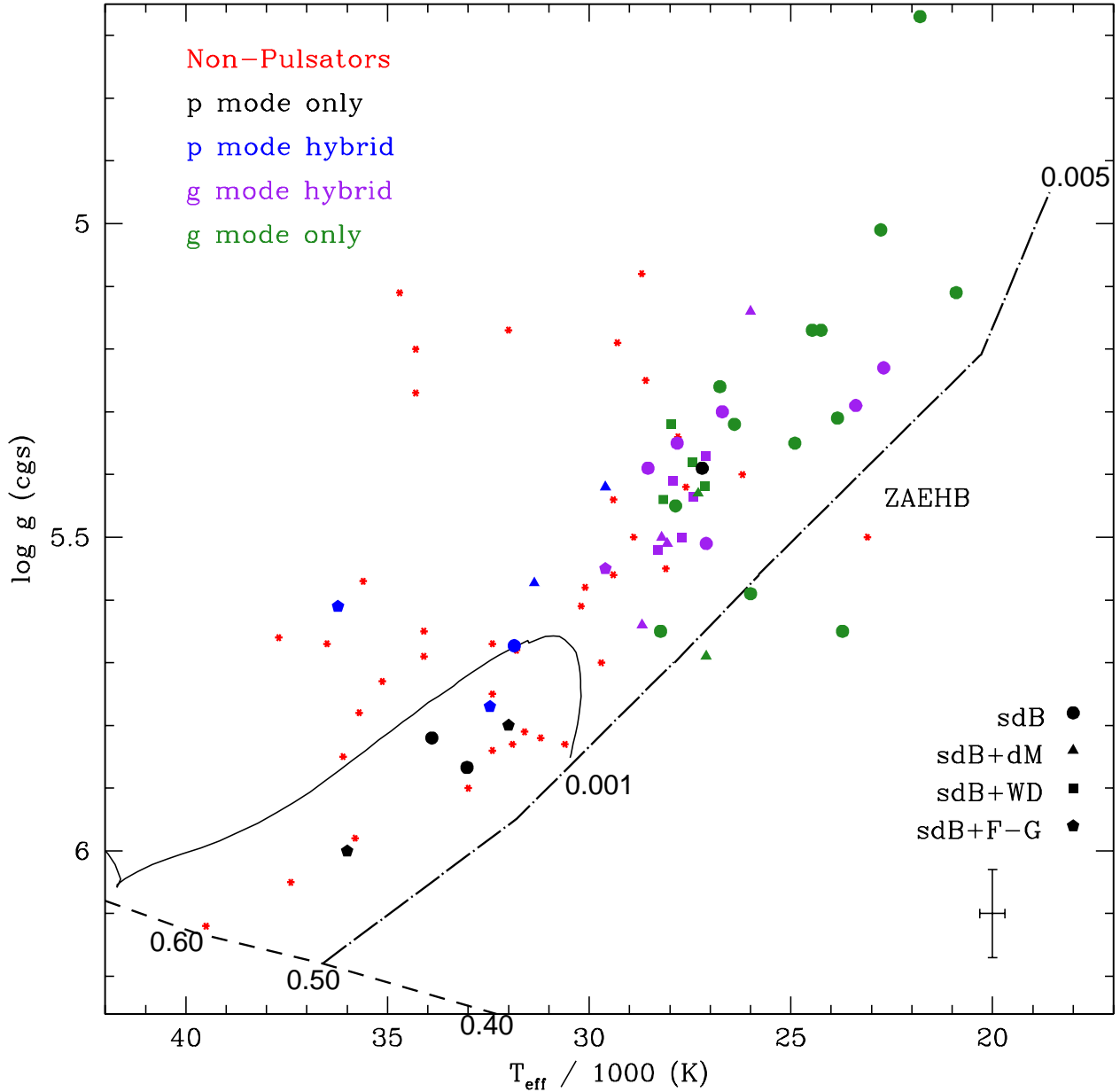,width=\textwidth} }
\caption{Kiel diagram of Kepler and TESS-observed sdB stars. Colours and symbols are indicated within
the figure: Non-pulsators are not shaped for binary information. 
The dashed-dotted line indicates the zero-age helium main-sequence (ZAEHB) for a core mass of $0.50M_{\odot}$
with varying envelope masses (two are labeled on the plot) and the dashed line indicates the ZAEHB for a
fixed envelope mass of $2\times10^{-4}M_{\odot}$ and varying core masses (three are labeled on the plot).
The solid lines shows evolution for a core mass of 0.5M$_{\odot}$ and an envelope mass of 0.001M$_{\odot}$.
Errorbars for the average of the errors from Table\,7 are indicated in the lower right.} %\ref{tabgrp1} are indicated in the lower right.}
\label{figHRK2C7}
\end{figure*}

Figure\,\ref{figHRK2C7} shows the locations of the pulsators in Table\,7 %\ref{tabgrp1}
in a Kiel diagram. Included in the figure are non-pulsators observed during K1 
\citep{roy10b,roy11b} and sample zero-age helium main-sequence (ZAEHB) and evolutionary
tracks \citep{reed04b}. It is well-known that
a larger fraction of sdB stars are observed to pulsate in g modes than p modes, or equivalently,
cooler sdB stars are more likely to pulsate at observable amplitudes
than hotter ones. Therefore it is not
too surprising that most of the Kepler-observed stars are g+p. Commensurate with
that is very few of the non-pulsators observed by Kepler are cooler sdB stars. There is only
one below 26\,000\,K while above 30\,000\,K 26 of the 28 (93\%) K1-observed stars are not
observed to pulsate.

\subsection{Hybrid pulsators}
\begin{figure*}
\centerline{\psfig{figure=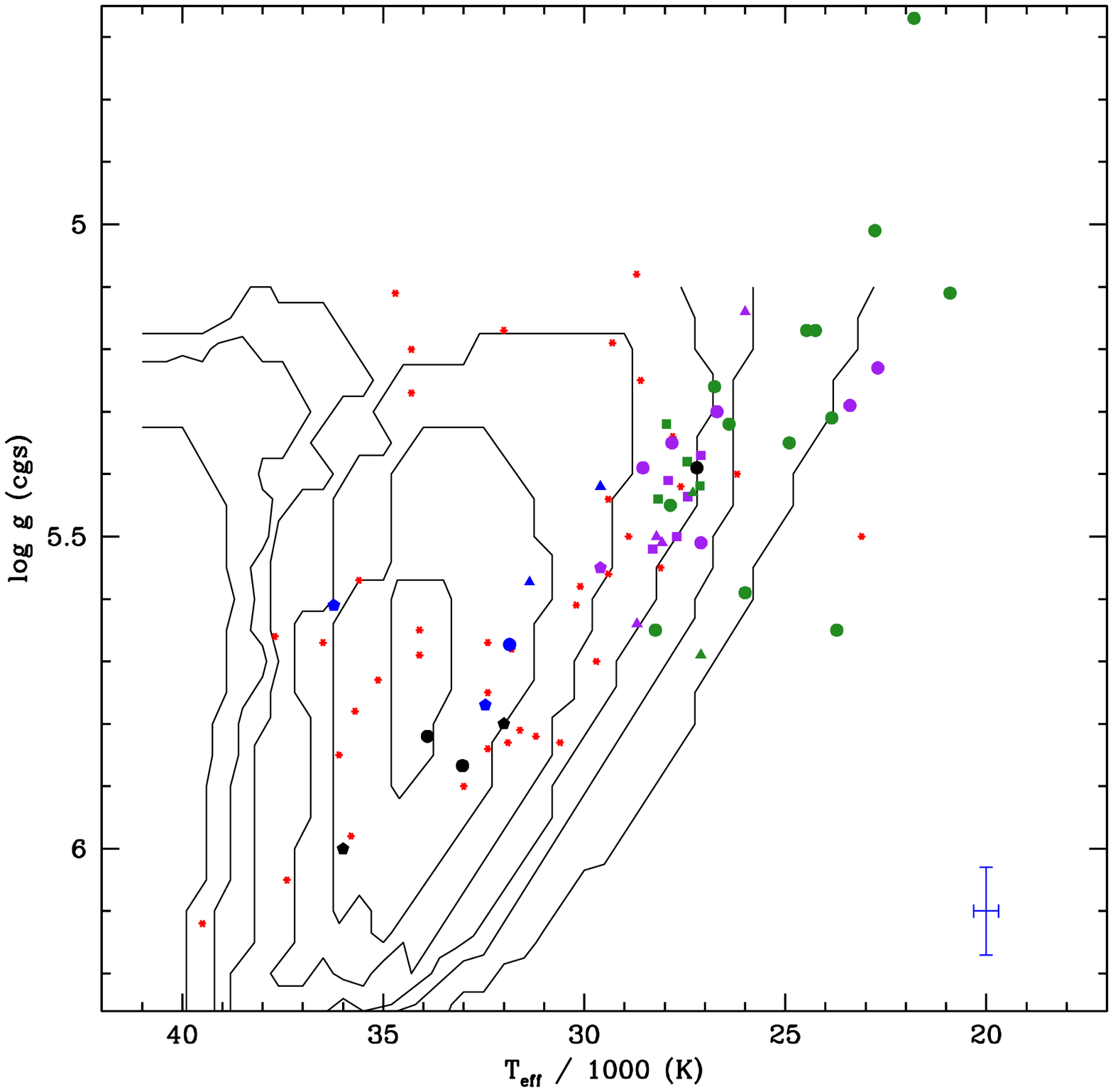,width=\textwidth} }
\caption{Kiel diagram showing p-mode instability contours \citep[reproduced
from][]{charpinet01} with points and average errorbars from Figure\,\ref{figHRK2C7}.}
\label{figHRcontour}
\end{figure*}

While models have largely been successful in predicting where p-mode pulsations
should occur in the Kiel diagram (contours in Fig.\,\ref{figHRcontour}), there
has been difficulty getting g-mode instabilities up to observed effective
temperatures \citep{jeffery06a,jeffery06b,hu09,bloemen14}. Prior to Kepler
observations, it was presumed that p-mode pulsations occurred in hotter sdB
stars, g-mode pulsations occurred in cooler sdB stars, and rare hybrid pulsators
would inhabit the temperature boundary. However, as can be seen in Figs.\,\ref{figHRK2C7}
and \ref{figHRcontour}, hybrid pulsators occur across nearly the full range of
temperatures, including the hottest and third coolest stars in our sample. 
In Tab.\,7 the transition from p+g to  g+p
%In Tab.\,\ref{tabgrp1}, the transition from p-mode-dominated to g-mode-dominated
occurs at 29\,000\,--\,30\,000\,K. The coolest p+g star and
the hottest g+p star both have $T_{\rm eff}\,=\,29\,600\,K$.

\subsection{Pulsators in binaries}
Table\,7 separates the pulsators by binary type.  All
%Table\,\ref{tabgrp1} separates the pulsators by binary type.  All
of our sdB binaries with white dwarf or M-dwarf companions are g+p,
though we know this is not a unique feature. Ground-based observations have observed
p mode sdBV stars with both types of companions. 
All save one of our Kepler/TESS-observed sdBV stars with F/G companions are p+g pulsators.
That sole g+p pulsator (CD$-$28$^{\circ}$\,1974), observed with TESS, is also
the hottest g+p star. 
Perhaps these trends are indicative of formation channels, as  the sdB+WD/dM binaries would
have experienced at least one common-envelope (CE) phase \citep{han02,han03} while the sdB+F/G binaries
would likely have had their envelopes stripped via Roche-lobe-overflow (RLOF) 
\citep[e.g.][]{vos18c}. There could also be an
observational bias in that cooler sdB stars are fainter than their F/G companions,
making them harder to detect. This could likely be answered by examining our 
full set (K1, K2, and TESS) of non-pulsators
which were initially selected using GALEX observations. If there were a reasonable
number of GALEX-selected sdB stars below 30\,000\,K with F/G companions, then we can
rule out an observational selection effect. This should be revealed in our forthcoming
K2 summary paper when pulsators and non-pulsators are compared.

Slightly more than half of our pulsators show no indication of a companion. This means
there were no indications of spectral lines from a companion, no RV
variations outside of standard deviations, and no photometric variations which could
be produced by the reflection effect, ellipsoidal variations, Doppler boosting, or
phase-induced pulsation aliasing caused by light-travel across an orbit. SdB formation
channels include stripping giant stars at the tip of the red giant branch via
RLOF or CE ejection, or by merging two helium white dwarfs (WDs). The result of the different
evolutionary paths is that the envelope-stripping ones produce sdB stars with masses
sharply peaked at the so-called ``canonical'' value of $0.47M_{\odot}$ or slightly
less while merged WDs can have a broad range of masses from 0.4 to over $0.7M_{\odot}$
\citep[See Fig.\,12 of][]{han03}. GAIA data should help answer this question as it
will detect astrometric binaries and, in many cases, determine masses when combined
with spectroscopic and/or spectral energy distribution data.

\begin{figure*}
\centerline{\psfig{figure=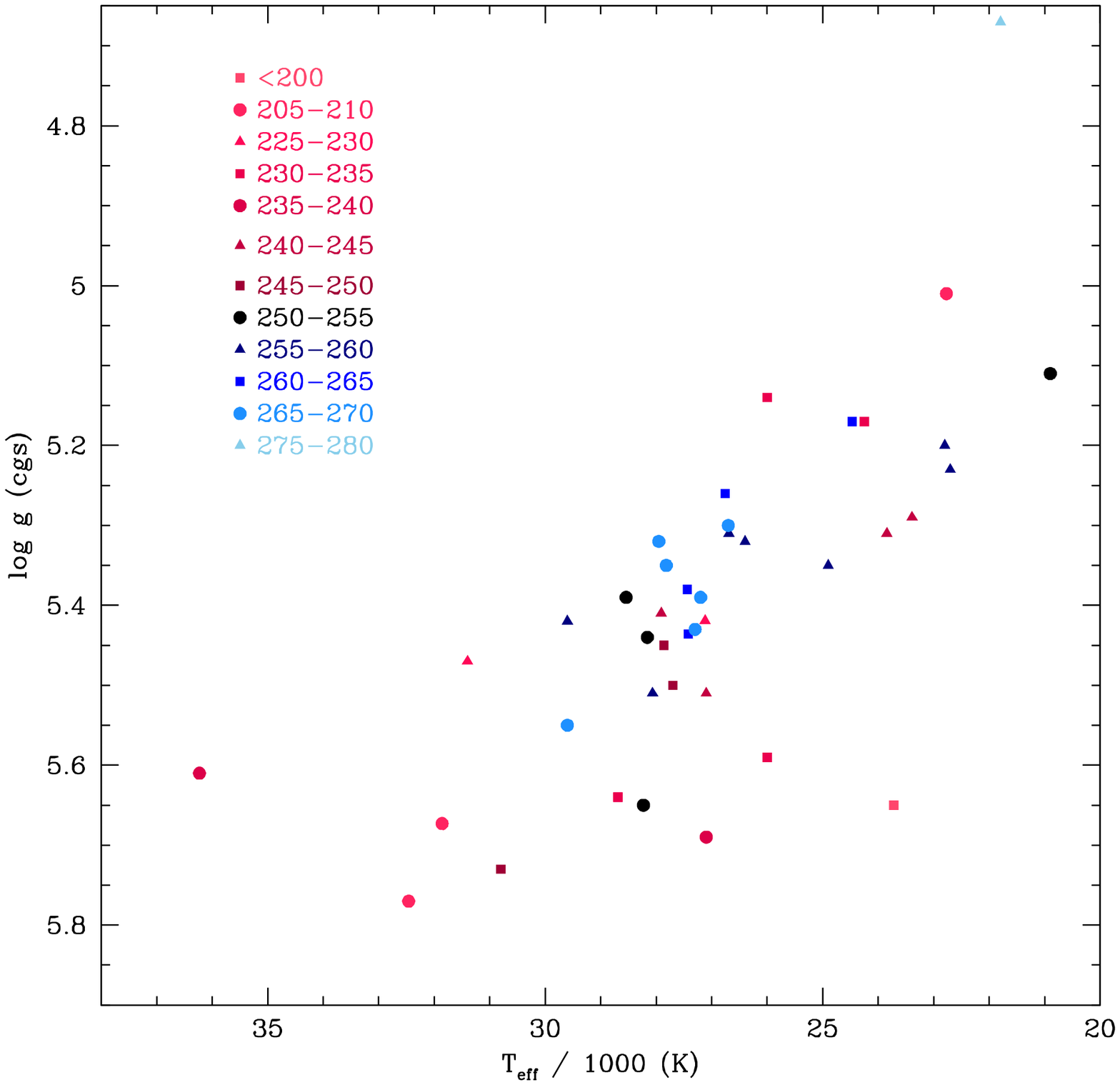,width=\textwidth} }
\caption{Kiel diagram showing $\ell =1$ g-mode period spacings, $\Pi_{\ell =1}$. Colour codings provided in
the figure and different point types are for clarity only.}
\label{figpspace}
\end{figure*}

\subsection{Observational correlations and trends}
Observations are meant to provide constraints and direction to models, from which
we determine the physics of stars. Observations we list in Table\,7 include %\ref{tabgrp1} include
spectroscopic ones, $T_{\rm eff}$, $\log g$ and binarity, and seismic ones, pulsation
type,
$P_{\rm Amax}$ (period of highest amplitude), $\Pi_{\ell =1}$ (g-mode
asymptotic period spacing for $\ell =1$), and rotation period (from frequency multiplets).

\begin{figure*}
%\centerline{\psfig{figure=freq_numax.eps,width=\textwidth} }
\centerline{\psfig{figure=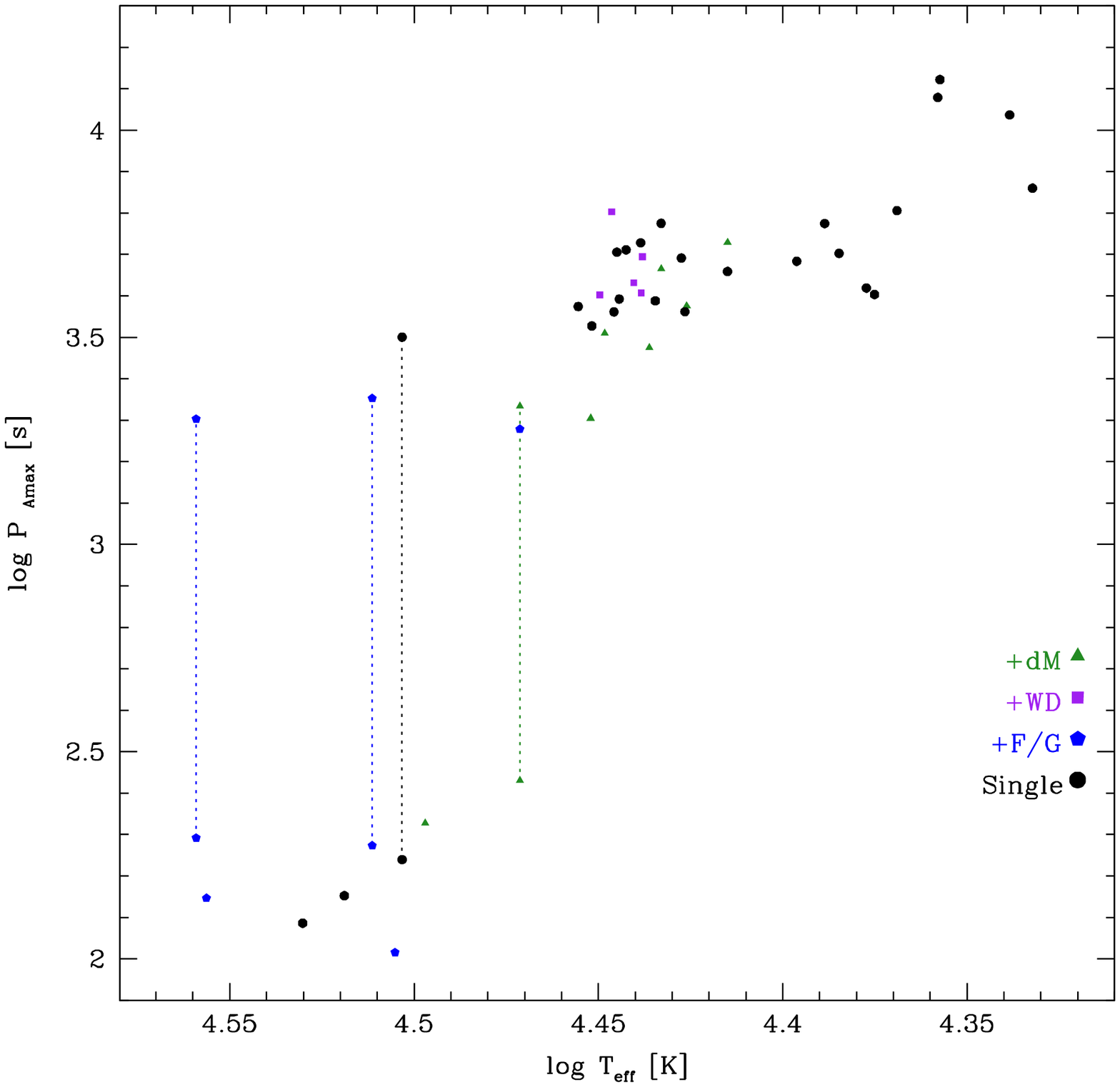,width=\textwidth} }
\caption{Comparing $T_{\rm eff}$ to $P_{\rm Amax}$. The four p+g pulsators have their
g-mode $P_{\rm Amax}$ connected to their p-mode $P_{\rm Amax}$ by dotted lines. Note
that CD$-$28$^{\circ}$\,1974 has had its point shifted down 0.02\,dex for clarity. 
Point shapes and colours (on-line only) are indicated in the figure.}
\label{figfreqrot}
\end{figure*}

Figure\,\ref{figpspace} is a Kiel diagram where
point colours indicate detected $\Pi_{\ell =1}$. There are no obvious trends.
The longer period spacings are mostly clustered in the middle (around
$T_{\rm eff}=27\,000\,$K and $\log g=5.35$), the median values (near $\Pi{\ell =1} =250-260\,s$)
extend from $T_{\rm eff}=23\,000\,$K to $T_{\rm eff}=30\,000\,$K, 
and the lower period spacings have tendencies to be towards
the extremes; $\log g>5.5$ or $T_{\rm eff}<27\,000\,K$, which also includes lower $\log g$.
Clearly asymptotic period spacings depend on something other than temperature or gravity.

Figure\,\ref{figfreqrot} shows the period of the
highest-amplitude pulsation ($P_{\rm Amax}$) in each star with $T_{\rm eff}$. Point
type and colour indicate binary information. As previously noted, the sdB+F/G binaries
are in the hottest group of stars. The p+g-mode pulsators show $P_{\rm Amax}$ for
both types of pulsations, connected by dotted lines.
Even though the $\log$ scale compresses the ordinate, there is a
clear trend of $P_{\rm Amax}$ towards longer periods in cooler stars.
This trend seems clear in the g/g+p region but less so in the p/p+g
stars. If we ignore the sdB+F/G stars, the p mode periods show a clear trend
in the same direction as the g modes, whereas if we include those, there is little correlation and even a
slightly reversed trend.

\begin{figure*}
\centerline{\psfig{figure=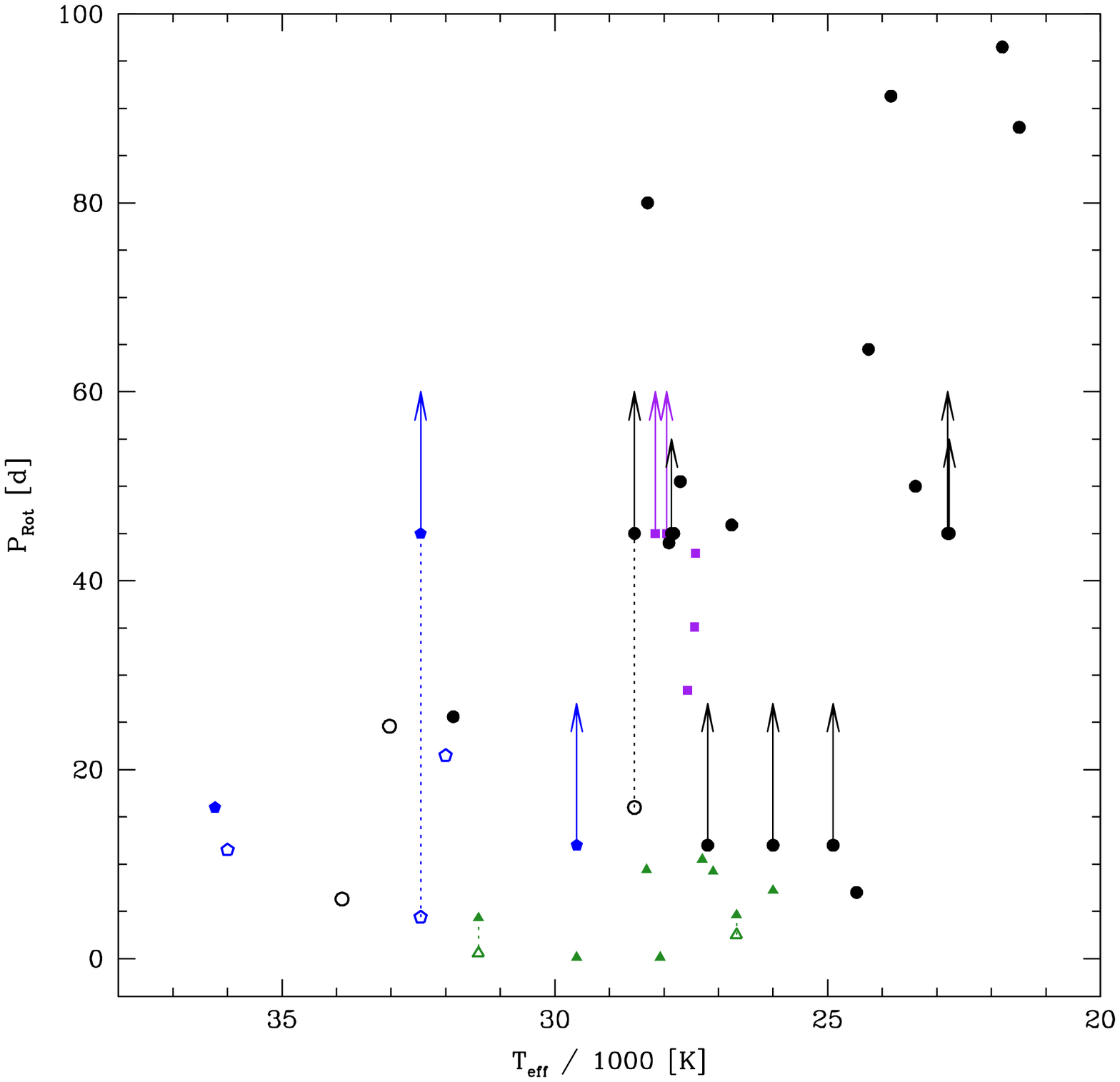,width=\textwidth} }
\caption{Comparing $T_{\rm eff}$ to rotation periods. Arrows indicate lower limits and dotted
lines connect rotation periods that are different for p and g modes. Open symbols indicate
rotation was determined from p-mode multiplets.
Point shapes and colours (on-line only) are the same as Fig.\,\ref{figfreqrot}.}
\label{figteffspin}
\end{figure*}

\subsection{Rotation}
It was pointed out in \citet{reed14} that there was a trend for cooler
stars to have longer rotation periods. With more pulsators studied, this is
re-examined in Fig.\,\ref{figteffspin}. There are four differential rotators (indicated
by points joined by a line) and K2 and TESS data often did not detect multiplets, which
means either the pulsations are pole-on, or we only have a lower limit on rotation. We presume
the latter, as shown by arrows in Fig.\,\ref{figteffspin}, as pole-on orientations are unlikely.
Two exceptions may be K10001893 and K8302197 for which no multiplets were detected in 
over 1\,000\,days of K1 data, indicating either a pole-on viewing angle, or extremely slow 
rotation ($>715$\,days).
Open points in Fig.\,\ref{figteffspin} indicate rotation was determined from p-mode multiplets.
It has been observed that in radially differential rotators \citep[][ and below]{mdr15} p modes
indicate faster rotation than g modes.
Rather than a correlation, we can now only state that below 24\,000\,K rotation is only slower
than 45\,days and above 32\,500\,K rotation is only faster than 25\,days. There are no binary
stars in our sample cooler than 26\,000\,K and no sdB+dM or sdB+WD binaries hotter than 32\,000\,K.
%It could be vaguely argued that there are two trends; one from cool and slower to hotter and
%faster, as originally mentioned in \citet{reed14} and the other for stars hotter than 24kK with slower rotation.
%The latter could be related to binarity, as many binary stars appear here, but there are
%exceptions to both trends.  However, there is a paucity of points in the slower--hotter quadrant,
%indicating that hotter stars tend to be the faster rotators. If the hotter stars have common
%core masses, the mechanism that produced their thinner envelopes could be related to their
%faster rotation.

\begin{figure}
\centerline{\psfig{figure=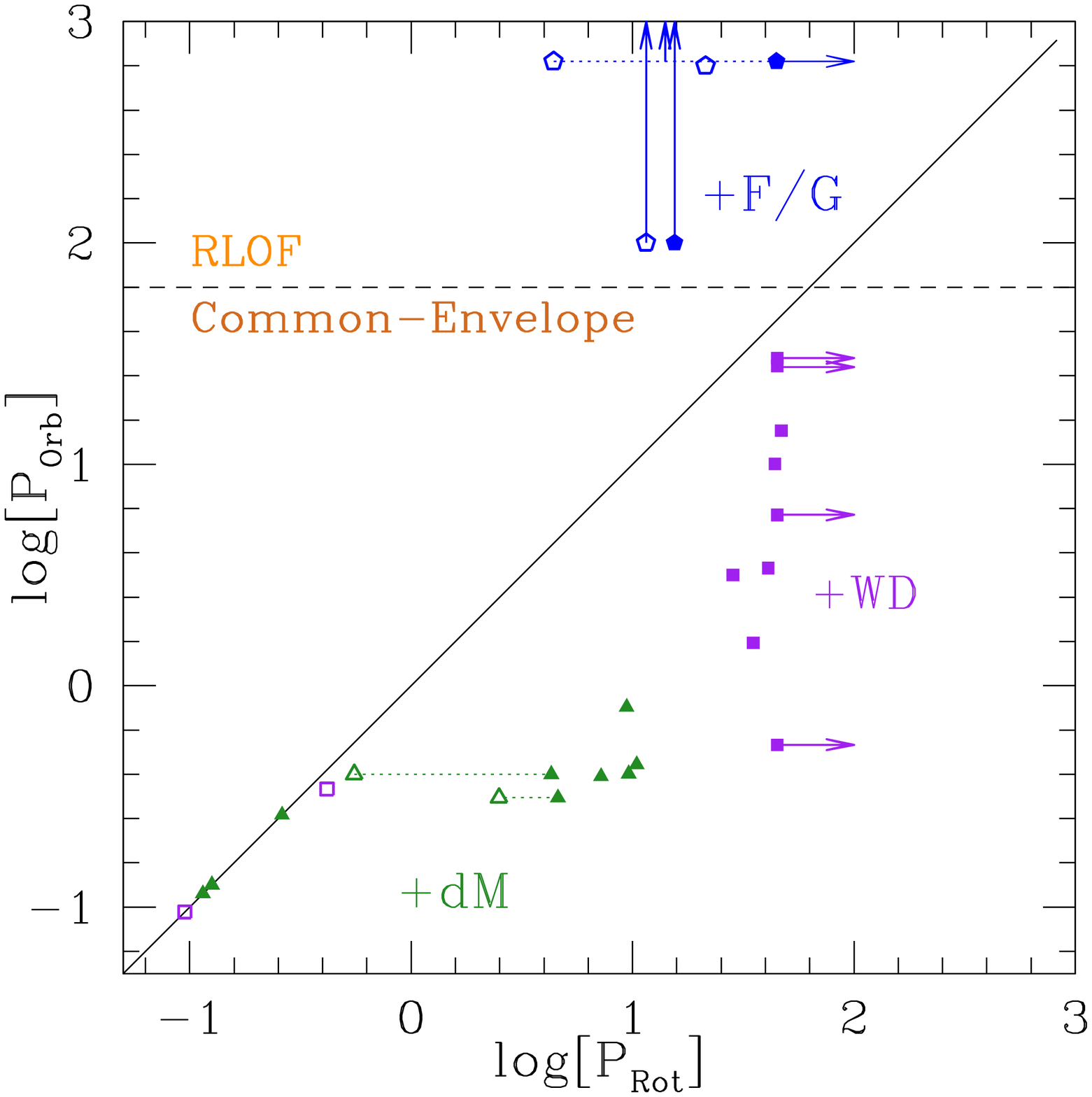,width=\columnwidth} }
\caption{Comparing orbital to rotation periods. Arrows indicate lower limits and dotted
lines connect rotation periods that are different for p and g modes. Open
points indicate rotation was determined from p-mode multiplets.
Point shapes and colours (on-line only) are the same as Fig.\,\ref{figfreqrot}.
Diagonal line indicates tidally-locked rotation.
The figure includes the sdB+WD binaries Feige\,48 \citep{reed12c} and
KPD\,1930$+$2752 \citep{mdr11} which have periods less than one day even
though neither was observed by Kepler. We also include the sdB+dM binary AA\,Dor
which is not a pulsator but has a spectroscopically determined rotation period \citep{vuc2016}.}
\label{figbinspin}
\end{figure}

The effect of binarity on rotation is investigated in Fig.\,\ref{figbinspin}.
The black diagonal
line indicates tidally-locked rotation.  
Interesting features in Fig.\,\ref{figbinspin} include that \emph{all}
space-based-observed sdB+WD stars are primarily g-mode pulsators that rotate subsynchronously.
As there are only eight such systems, and g-mode pulsations occur more often than p-mode ones,
this may be a selection effect. We include the two p-mode sdB+WD
stars Feige\,48 \citep{reed12c} and KPD\,1930$+$2752 \citep{mdr11} (from ground-based data) 
as they are the shortest-period sdB+WD binaries which also have frequency multiplets to indicate
rotation period.  KPD\,1930$+$2752 is tidally locked while Feige\,48 is nearly so. All the
g-mode sdB+WD binaries have very similar rotation periods, though four of those only have lower limits on rotation. 
The sdB+dM binaries all have
short binary periods, and this is almost certainly a selection effect. They are usually
detected by the so-called ``reflection effect'' where the sdB stars heat one side of the dM
stars, causing brightness variations with the orbital period. 
This effect will not be detected if the binary separation is too large and the 
dM will not be observed as the sdB star far outshines it \citep[e.g. \S 3 of][]{reed04}.
 The sdB+dM binaries in our sample are
tidally locked until the binary period exceeds $\sim$0.25\,d then they all rotate subsynchronously.
The sdB+F/G stars in our sample rotate commensurate with their apparently single sdB counterparts
in the same temperature range (Fig.\,\ref{figteffspin}) but supersynchronous to their long-period
orbits.

\begin{figure}
\centerline{\psfig{figure=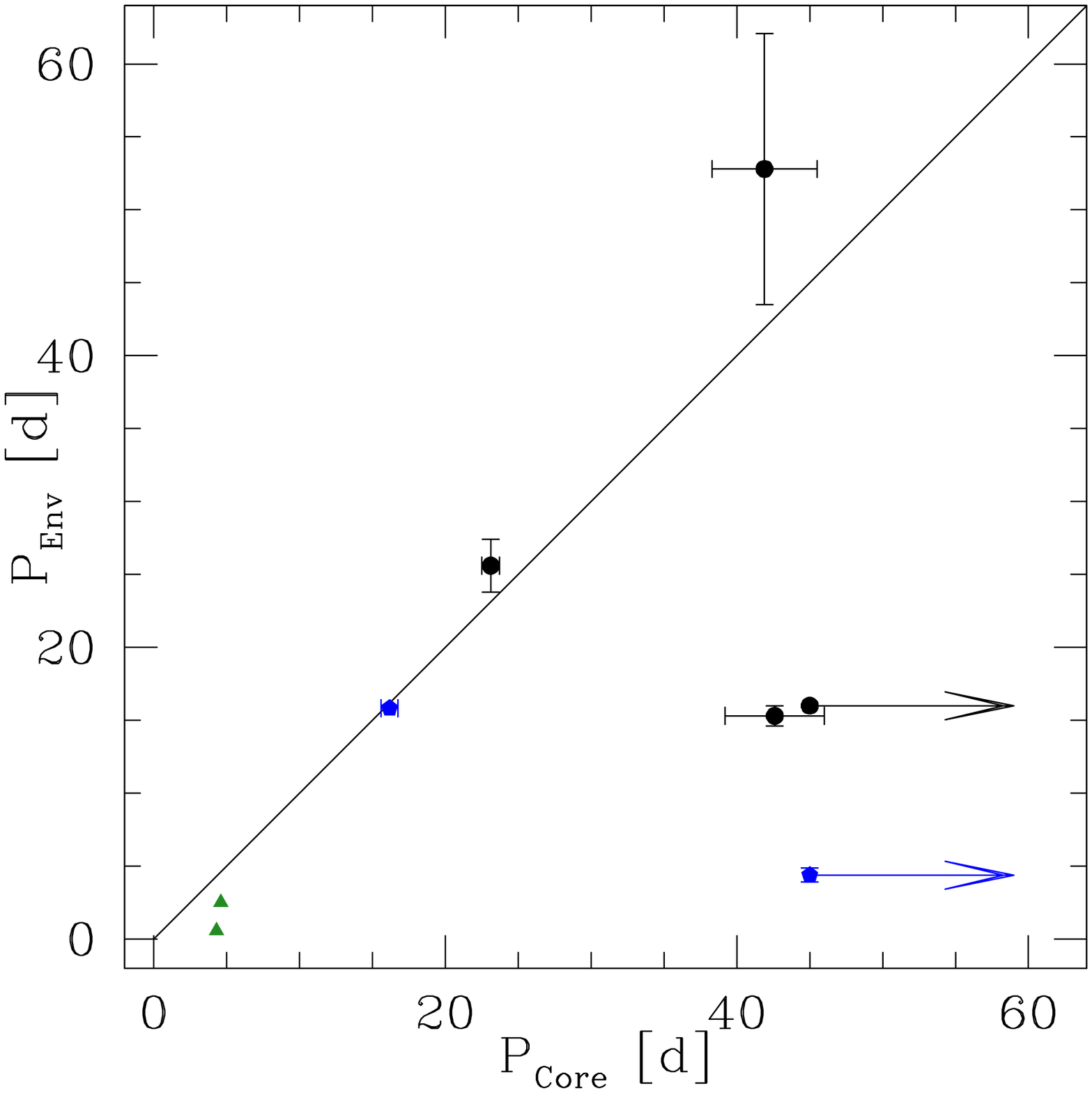,width=\columnwidth} }
\caption{Comparing envelope to interior rotation periods for stars where multiplets
have been detected in both g and p modes. PG\,0048 and E211779126 only have lower
limits for their g-mode pulsations. Errorbars are omitted when 
smaller than the points. Point shapes and colours (on-line only) are
the same as Fig.\,\ref{figfreqrot}.}
\label{figradspin}
\end{figure}

%\subsection{Radially differential rotators... or not}
Figure\,\ref{figradspin} shows the eight stars where multiplets have been 
detected in both p and g modes, or there were sufficient pulsations that
they should have been, providing a lower limit. It is expected that p-mode pulsations
mostly sample the envelope while g-mode pulsations sample deeper into the star
\citep{charp14}.
To date there have been five that appear to be radially-differential rotators % 4 5 20 2 29
\citep{baran16d,mdr15,reed19a,mdr19,andy19} and three that likely
rotate as solid-bodies \citep{baran12a,kern16,reed19a}. % 3 11 20 
There are no obvious correlations with binarity as half of the apparently single and sdB+F/G stars
rotate differentially and half rotate like solid bodies. 
Both of the sdB stars with dM companions rotate differentially.
In all cases of differential rotation, p-mode pulsations indicate faster
rotation than g-mode pulsations, with the possible exception of KIC\,10139564 \citep{baran12a}
which has $1\sigma$ errorbars that overlap the solid-body rotation line.

\section{Discussion}
\subsection{Campaign 7 Pulsators}
In this paper we analyzed four newly-discovered sdBV stars and add them to the growing
number of sdBV stars with a large fraction of identified pulsation modes.  We detect
asymptotic g-mode period spacing sequences in all four, with three having $\Pi_{\ell =1}$ values
right near 250\,s and the other near to 200\,s. Only three other sdBV stars have such
unusually short asymptotic spacings; two are hot p+g pulsators and the
other another cool g-mode-only pulsator. So the extremely low asymptotic spacings
only occur for stars at both extremes. However, stars with similar physical properties
to each group also have ``normal'' spacings near to 250\,s and so this mystery has yet
to be solved. 

Three of the stars in C7 are apparently single, while we have discovered \ethree\, to
have RV variations, indicative of a white dwarf companion in a 5.92\,d orbit. Somewhat
surprisingly we do not detect any multiplets in \ethree\, and so presume the rotation period
to be $>45$\,d. Only in \efour\, do we detect multiplets which indicate a rotation period
near 7\,d. We could have anticipated detecting two sdBV stars with multiplets based on the
56\% detection rate from K1 for periods under 45\,days.

\subsection{The Group of Kepler and TESS-observed Pulsating Subdwarf B Stars}
%Adding these four stars to those already solved using Kepler or TESS data provides several tests
%which can be done to search for features indicative of physical properties. 
We completed ensemble analyses so that modelers may use this information to deduce
physical processes within these stars. In our analyses we have only included Kepler and TESS-observed
sdBV stars as it provides a roughly homogeneous observing sample with the highest quality
data.
In examining the group, we think we have revealed some underlying relationships which may prove fruitful
for model study.

\subsubsection{Pulsations}
As has been known for some time, pulsators and
non-pulsators overlap in the Kiel diagram (Figs.\,\ref{figHRK2C7} and \ref{figHRcontour}), 
with a higher percentage of cool sdBV stars
pulsating with g modes. It has been observed that hotter stars favour p modes and
cooler ones favour g modes, and we also find this. Of the 43 stars in our sample, the dividing
line between p+g and g+p pulsators is 29\,000-30\,000\,K.
The p/p+g pulsators cluster around the central instability contour indicating
that pulsation driving models are likely reasonably accurate. Driving models for g-mode pulsations
have been less certain (which is why instability contours are not included in 
Fig.\,\ref{figHRcontour}) as the temperature range seems strongly dependent on the
amount of enhancement of iron-group elements in the driving region \citep[e.g.][]{jeffery06b,hu09}. 
More problematic are the hybrid pulsators. Originally discovered in a limited mid-temperature range bordering
p and g pulsations, Kepler and TESS observations detect them from 22\,800 to 36\,300\,K.
49\% of the stars in our sample are hybrid pulsators. Pulsation driving models will need to account for the
two hybrids with p modes completely outside
those instability contours and g-mode pulsations that span from 22\,800 to 36\,300\,K. %Driving models will need
%to account for these extended ranges of pulsations.

Our sample has seven stars (16\%) with $T_{\rm eff}$ below 25\,000\,K and none are in known binaries. 
Similarly there are no PCE (sdB+WD/dM) binaries with $T_{\rm eff}$ above 31\,500\,K in
our sample (also seven stars), though at least two are known from ground-based observations; the
sdB+dM binary PG\,1336-018 \citep[][]{kilkenny98} and the sdB+WD binary KPD\,1930+2752 \citep{billeres00}. 
This poses two questions; Does binarity shut off pulsation
driving below 25\,000\,K? And does PCE binarity adversely effect pulsation driving
above 31\,500\,K? It is likely too early to draw conclusions (particularly on the hot side,
where two systems are known) but this will be worth
watching as the sample is completed for K2-discovered pulsators and increased with TESS data.

Of the relationships we examined, the one with 
the clearest correlation appears to be between $T_{\rm eff}$ and P$_{\rm Amax}$ which shows a clear
trend for cooler stars to have longer periods (lower frequencies) of their highest-amplitude  pulsations.
\citet{shoaf20} noted two g-mode
outliers below the trend that both have indications of smaller-than-canonical masses. If the
$T_{\rm eff}$--P$_{\rm Amax}$ relationship is indicative of the resonant cavity size, then stars
along this relationship would likely have a common convective core size (inner boundary of
the resonant cavity) with varying envelope thicknesses. Thicker envelopes equate to lower
$T_{\rm eff}$, a larger resonant cavity, and therefore longer P$_{\rm Amax}$.  Another outlier
(E248368658) above the trend could indicate a higher-massed core.

Related to the $T_{\rm eff}$--P$_{\rm Amax}$ relationship, we anticipated finding a similar 
relationship with mean period spacings, $\Pi_{\ell =1}$, but do not. A recent study by \citet{uzu21}
examined $\Pi_{\ell =1}$ in models with a narrow range of total
mass and two convective core masses. They note that $\Pi_{\ell =1}\propto \bar{g}^{-1}R_{\star}/
\left(R_{\star}-\,R_{core}\right)$. From this, they found agreement with previous studies
\citep[e.g.][]{cast85,constantino15,ost21}, that smaller convective cores have smaller $\Pi_{\ell =1}$
and stars with larger convective  cores, and consequently lower $\log g$,
have larger $\Pi_{\ell =1}$. Those results could be
indicative of why Fig.\,\ref{figpspace} shows no obvious patterns.

\subsubsection{Rotation}
We examined the correlation
between rotation period and $T_{\rm eff}$ (Fig.\,\ref{figteffspin}), as noted by \citet{reed14}, and find that it
only seems to hold at extreme values of $T_{\rm eff}$. In our sample there are no fast-rotating
cool nor slowly-rotating hot sdBV stars with no known binaries at 
all in our cool sample
and no PCE (sdB+WD/dM) binaries in our hot sample. 
The mid-$T_{\rm eff}$ star E217280630 has a rotation period of 7\,days
which is in a range dominated by sdB+WD binaries. It would be worth obtaining additional observations 
to search for a white dwarf companion. 

Unfortunately there is currently little hope for additional
constraints on those stars with only lower limits on their rotation periods. No ground-based observations
have resolved multiplets in g-modes of sdBV stars and transparency variations, which occur on
similar time scales to those pulsation periods, make it extremely difficult even to observe
g-mode sdBV pulsations without space-bourne telescopes. As K1 observations show, only long-duration 
space-based observations are likely to fully resolve frequency multiplets and provide rotation periods.

Patterns do emerge when comparing orbital to rotation periods of known binary systems. First, all PCE
stars in our sample with binary periods longer than 0.3\,days rotate subsynchronously while those
shorter than 0.3\,days are synchronized. 

All of our
sdB+WD systems with measured rotation periods are between 28 and 51\,days. There are two known
sdB+WD systems with shorter binary periods and also rotation periods under one day (Fig\,\ref{figbinspin}).
While 44\% of our sample only have lower limits near 45\,days, we would speculate that sdB+WD systems,
which have undergone two common-envelope phases, have a common PCE initial rotation period near 45\,days
which then evolves towards shorter periods. 

We see a very similar pattern in our sample of sdB+dM binaries.
Four stars have periods near 10\,days, three stars are tidally locked with periods under 0.3\,days, and
two are in between and differentially rotating with the envelope spinning faster. This has two
possibilities; either the rotation of these binaries, which have only had one common-envelope epoch, 
is correlated with the binary period, or these PCE systems also have a common initial rotation period
which evolves to shorter periods. We think the latter is the most likely explanation, particularly since
the differential rotators have faster-spinning envelopes which are likely being spun up by the companions,
through tidal interactions. Furthermore, the dissipation of tidal energy can be achieved in the
form of pulsations.

Our sample of sdB+F/G stars are only on the hot end and rotate faster than the average but not
dissimilar to single sdB stars in the same temperature range. It is presumed their temperatures
are related to thinner envelopes yet material deeper within (g mode) sdB stars 
is correlated with slower rotation. As these stars likely formed via RLOF perhaps that mechanism
produces faster rotators, or conversely, RLOF does not slow rotation during mass loss like the
CE mechanism does. A larger sample of sdB+F/G stars may answer this question.

It is more difficult to find meaning with differential rotation (Fig.\,\ref{figradspin}). Both of
the sdB stars with dM companions for which we have both p and g-mode multiplets have differential 
rotation. It is unlikely a common property of all sdB stars with dM companions as two sdB stars with dM
companions in Fig.\,\ref{figbinspin} have g-mode multiplets indicating they are tidally locked. As all
the sdB+dM rotation periods are short, TESS data will likely expand this sample.
Our
sample of sdB+F/G stars and those without known companions shows a mixture. We would encourage investigations
to determine if any of these systems have short-period companions, as difficult as that may be. Only by
ruling out binarity with periods up to ten days can we determine if differential rotation is inherent
to a property unrelated to binarity. If it is not related to binarity, then a companion ``spinning up''
sdB stars cannot be the cause and it must therefore be related to mass loss near the tip of the red
giant branch.

The above relationships likely relate to mass loss and  angular momentum transport
during their formation mechanisms. Another piece of evidence is the comparison between
binary and rotation periods. Most PCE sdB stars in our sample rotate subsynchronously to their orbit unless it is
less than about half a day. This information should be useful in modeling PCE binaries.

\subsection{The Future}
Our examination of the group properties includes an intermediate sample from K2 and a very preliminary
sample from TESS. There are over a dozen suspected pulsators observed during K2 which have yet to
be thoroughly analyzed and
TESS has now observed thousands of sdB stars, with perhaps a couple hundred new pulsators. Those data
should produce a statistically significant sample from which to examine pulsation properties. However,
even with an expanded sample, rotation is likely to be problematic as K1 showed us that we really need
continuous data for more than a year before we are likely to resolve frequency multiplets.

We did not examine individual periods of the stars in our ensemble in our analyses. So in 
addition to the properties discussed in this paper, we encourage modelers to examine properties which
include trapped modes, radial indices where asymptotic sequences begin and end, the ``hook' feature in
asymptotic sequences, and the standard deviation in $\Pi$ \citep[e.g.][]{constantino15}. Additionally,
it may well be worth a look at other ``hook''-feature pulsators to see if they too have a different
linear asymptotic period sequence shortward of the bend, as \ethree\, does.

Separate from pulsations, GAIA is providing reliable parallaxes \citep{BJ2018DR2,gaiadr3}, from which distances, 
radii, and masses can be 
determined \citep[e.g.][]{kilkenny19,andy19,mdr19}. In combination with the remaining K2 sdBV stars, and
the many yet to be discovered by TESS, powerful observational tools are becoming available. 
We are likely on the brink of understanding
the underlying relationships hinted at in this paper.
\smallskip

{\bf Data Availability Statement:} The data underlying this article are available in the
Mikulski Archive for Space Telescopes (MAST). \url{https://archive.stsci.edu/}
Data obtained with the Nordic Optical Telescope is available after 
a one-year proprietary period. \url{http://www.not.iac.es/archive/}
\smallskip

ACKNOWLEDGMENTS: Funding for this research was provided by the National Science
Foundation
grant \#1312869 and NASA grant 14-K2GO2-2-0047 as part of the K2 guest observer
program. Any opinions, findings, and conclusions or
recommendations expressed in this material are those of the
authors and do not necessarily reflect the views of the National
Science Foundation or NASA. JAC was funded by the Missouri Space Grant, which
is funded by NASA.
ASB gratefully acknowledges financial support from the Polish National Science
Centre under projects No.\,UMO-2017/26/E/ST9/00703 and NO\. UMO-2017/25/B/ST9/02218.
This paper includes data obtained by the \emph{Kepler} mission. Funding for the
 \emph{Kepler} mission is provided by the NASA Science Mission directorate.
Data presented in this paper were obtained from the Mikulski Archive for
Space Telescopes (MAST). STScI is operated by the Association of Universities
for Research in Astronomy, Inc., under NASA contract NAS5-26555. Support
for MAST for non-HST data is provided by the NASA Office of Space Science via
grant NNX13AC07G and by other grants and contracts.

The spectroscopic observations used in this work were obtained with the
Nordic Optical Telescope at the Observatorio del Roque de los Muchachos
and operated jointly by Denmark, Finland, Iceland, Norway, and
Sweden.

\bibliography{sdbrefsMNRAS}

\begin{landscape}
\begin{table}
\label{tabgrp1}
\centering
\caption{Spectroscopic and pulsation properties of published Kepler and TESS-observed
sdBV stars organized by binary type, pulsation type, and then decreasing effective temperature.
Column 1 provides the KIC, EPIC identifications, column 2 other
identifications, column 3 provides the binary status where sdB is listed for single stars,
sdB+WD for those with white dwarf companions, sdB+dM for those
with M-dwarf main sequence companions and sdB+F or sdB+G for
those with main sequence F or G companions. Column 4 provides the
pulsation type where p are p-mode only pulsators, p+g are predominantly
p mode hybrid pulsators, g are g-mode only pulsators, and g+p are
predominantly g mode hybrid pulsators. Column 5 lists the g-mode $\ell =1$ asymptotic period spacing,
Column 6 the maximum
amplitude and Column 7 the pulsation period of maximum amplitude. Columns 8 and 9 provide spectroscopic
properties. Column 10 provides the binary period, column 11 provides the rotation
rate and if two values are given, the first is derived from p-mode multiplets (for the envelope) and the
second from g-mode multiplets (for deeper interior), and column
12 has references. The references are 1: \citet{roy14b}, 2: \citet{mdr19} 3: \citet{baran12a},
4: \citet{baran16d}, 5: \citet{mdr15}, 6: \citet{sahoo20}, 7: \citet{charpinet11b},
8: \citet{reed11c}, 9: \citet{uzu}, 10: Kern thesis 11: \citet{kern16}, 12: \citet{silv19},
13: \citet{bachu16}, 14: this work, 15: \citet{andy15b}, 16: \citet{ketzerF2}, 17: \citet{charp19}, 18: \citet{reed14},
19: \citet{F5_paper}, 20: \citet{reed19a}, 21: \citet{shoaf20},
22: \citet{telting12a,kern18}, 23: \citet{jht14}, 24: \citet{roy14a}, 25: \citet{ReedPG1142},
26: \citet{andy16}, 27: \citet{2m1938}, 28: \citet{andy19}, 29: \citet{baran18},
30: \citet{pablo12}, 31: \citet{baran12c}, 32: \citet{ostensen12b}, 33: \citet{csj17}
Notes: $^{\diamondsuit}$ indicates stars with pulsation results only from the one month survey data.
$^{\star}$ No multiplets were detected
for these stars which means spin periods longer than the observations, or that the pulsation axis is
very nearly pole-on. In these cases, a lower limit of the spin period is provided. $^{\dagger}$ RV
excesses have been measured for these stars (Sanjayan, S. et al. \emph{in press}). K1718290 is listed as a blue horizontal
branch (BHB) star. }

\begin{tabular}{lccccccccccccl}
Kepler ID & Other & Binary & Type &  $\Pi_{\ell =1}$ & $A_{\rm max}$ & $P_{\rm Amax}$ &$T_{\rm eff}$ & $\log g$ & P$_{\rm orbit}$ & P$_{\rm spin}$ & Ref.\\
          &       & Status &  & (s) & (ppt) & ($\mu$Hz)   & (K/1\,000) & (dex, cgs) & (d) & (d) & \\ \hline
K2991276 &  & sdB & p &--  & 2.25  & 8201.2             & 33.9 (2) & 5.82 (4) & -- & 6.3 &  1\\
E248411044 & UY\,Sex & sdB & p&  -- & 4.89 & 7038.163           & 33.03 (20) & 5.88 (1) & -- & 24.6 (3.5) & 2 \\
K10139564 &  &  sdB & p+g  & 207/310  &7.95   &5760.2           & 31.86 (0.13) & 5.67 (3) &-- &  25.6 (1.8)/23.12 (62) & 3\\
E211779126 & 2M\,0856+1701  & sdB & g+p & 256 & 0.795  & 266.5 & 28.54 (8) & 5.39 (1) &  -- & $16/>45^{\star}$ & 4\\
K3527751 & & sdB & g+p & 266.4 (2)  & 7.11  & 255.7             & 27.82 (16) & 5.35 (3) & -- & 15.3 (7)/42.6 (3.4) &  5\\
T169285097  & SB\,815 & sdB & g+p & 265.09 (6) & 1.642 (21) & 258.1879 (29) & 27.20 (55) & 5.39 (10) & -- &  $>12^{\star}$ & 6 \\
K5807616$^{\diamondsuit}$ & KPD1943+4058  & sdB & g+p & 242.12 (62)  & 0.142  & 167.8   & 27.1          & 5.51        & -- &  -- &7,8\\
K10001893 &  & sdB & g+p & 268.0(5)  & 1.162  & 274.3           & 26.7 &        5.3         & -- &  $>715^{\star}$ &9\\
K2569576& NGC\,6791\,B3 & $^{\dagger}$sdB & g+p & 252.27 (66)  & 3.601  & 198.4         & 24.25 (46) & 5.17 (5) & -- &  64.5 (8.2) & 10\\
K2697388 &  & sdB & g+p  & 240.06 (19)  & 33.29  & 156.4                & 23.39 (12) & 5.29 (2) & -- & 41.9 (3.6)/52.8 (9.3) &  11\\
E220641886 & HD\,4539  & sdB & g+p & 256.5 & 0.80 & 83.4        & 22.80 (16) & 5.20 (2) & -- &$>45^{\star}$ & 12 \\
E212707862 & & sdB & g  & 252.6 (1.1) & 0.434 & 296.9           & 28.30 (16) & 5.48 (3) & -- & 80 &  13\\
E215776487 &  & sdB & g  & 247.8 (1.5)  & 0.489  & 197.0                & 27.86 (16) & 5.45 (2) &  -- & -- & 14\\
K8302197 &     & sdB & g & 258.61 (62)  & 0.87  & 187.0                 & 27.45 (20) & 5.44 (3) & -- & $>715^{\star}$ &  15\\
E203948264 &  & sdB & g  &261.3 (1.1)  & 0.722  & 203.5         & 26.76 (61) & 5.26 (9) & -- & 45.9 (8) &  16\\
T457168745 & PG0342+026 & sdB & g  & 232.25 (30) & 0.819 (21) & 219.274 (6) & 26.0 (1.1) & 5.59 (12) & -- & $>12^{\star}$ & 6 \\
T67584818   & SB\,459 & sdB & g          & 259.16 (56) & 1.72 (5) & 207.314 (9) & 24.9 (5) & 5.35 (10) & -- & $>12^{\star}$ & 6 \\
E218717602&  & sdB & g  & 263.55 (61)  & 3.71  &168.1           & 24.47 (16) & 5.17 (2) & -- & -- & 14\\
K2437937 & NGC\,6791\,B5 & $^{\dagger}$sdB & g & 248.9 (1.3)  & 1.03  & 240.5   & 23.84 (68) & 5.31 (9) & -- &  91.3 (14.1) &10\\
T278659026 & EC\,21494-7018 & sdB & g &  196.8 & 0.3184 (43) & 249.269 (3)  & 23.72 (23) & 5.65 (3) & -- &  $>12^{\star}$ & 17 \\
E217280630 &  & sdB & g  & 207.35 (21)/195.28 (39)  & 0.39  & 75.5              & 22.77 (15) & 5.01 (2) & -- & 7 & 14\\
K10670103 &  & sdB & g  & 251.6 (2)  & 13.99  & 138.1           & 21.49 (54) & 5.14 (5)  & -- & 88 (8) &  18\\
K1718290 &  & BHB & g   & 276.3 (1) & .268  & 91.9      & 21.80 (14) & 4.67 (3) & --& 96.5 & 33\\ \hline

E211823779 &  & sdB+F1V& p & --  & 2.052  & 7131.9538 (5)               & 36 & 6 & -- & 11.5 (8) &  19\\
E211938328 & LB\,378, EGGR\,266  & sdB+F6V      & p & -- & 2.00  & 9648.77 &32 & 5.8 & 635 (146) & 21.5 (6) &  18\\
E212508753 & PG\,1315-123  & sdB+G & p+g & 236.5 (1.3)  & 1.79  & 8116.1/497.62         & 36.23 (71) & 5.61 (9) &  $>100$ & 15.83 (19)/16.18(57) &  20\\
E220614972& PG\,0048+091 & sdB+G & p+g  & 207.45 (40) & 1.792 & 5339.2/443.7    & 32.46 (27) & 5.77 (6) & $>100$ & 4.39 (48)/$>45^{\star}$ & 20 \\
T13145616 & CD$-$28$^{\circ}$\,1974 &  sdB+F/G & g+p & 268.85 (32) & 1.894 (91) & 469.215 (12) & 29.60 (38) & 5.55 (9) & $>1\,000$ & $>12^{\star}$ & 21 \\ \hline

%E248368658 &  & sdB+WD & g+p & 245  & 1.21  & 152.4            & 30.8 & 5.73 & 30.2 & $>45^{\star}$ & 22 \\
K11558725 &   & sdB+WD & g+p    & 244.45 (32)  & 0.95  & 274.64187 (2)          &  27.91 (32) & 5.41 (1) &10.055 (5) & 44 &  22\\
K7668647 & FBS1903+432 & sdB+WD & g+p & 248  & 0.42  & 194.5    & 27.7 (3)  & 5.50 (3)  & 14.174 (4) & 50.5 (5) & 23\\
K10553698 &  & sdB+WD & g+p & 263.15  & 1.367  & 202.0                  & 27.42 (0.29) & 5.44 (24)&  3.387 (14) & 42.9 &  24\\

E218366972 &  & sdB+WD & g  & 254.95 (50) & 2.742  & 249.8      &  28.16 (11) & 5.44 (2) & 5.92 (1) & $>45^{\star}$ & 14\\
E201206621 & PG\,1142-037 & sdB+WD & g & 267.9 (10)  & 0.34  & 137.38   & 27.954 (54) & 5.32 (1) & 0.54109 (2) & $>45^{\star}$ &  25\\
E211696659 & & sdB+WD & g & 227.05 (56) & 0.188 & 233.41                        & 27.57 (30) & 5.70 (3) & 3.1604 (15) & 28.4 (1.4) &  17\\
K7664467 &   & sdB+WD & g & 263  & 0.495  &247.0                        & 27.44 (12 ) & 5.38 (2) & 1.5591 (6) &35.1 (6) &  26\\ \hline
\end{tabular}
\end{table}
\end{landscape}

\begin{landscape}
\begin{table}
\contcaption{}
\begin{tabular}{lccccccccccccl}
Kepler ID & Other & Binary & Type &  $\Pi_{\ell =1}$ & $A_{\rm max}$ & $P_{\rm Amax}$ &$T_{\rm eff}$ & $\log g$ & P$_{\rm orbit}$ & P$_{\rm spin}$ & Ref.\\
E246683636 & V1405\,Ori & sdB+dM & p+g & 225 & 4.534 & 4703.538         & 31.4 (2) & 5.47 (4) &  0.398023 (0.3) & 0.555 (29)/4.3 & 2 \\
K9472174$^{\diamondsuit}$ & 2M1938+4603  & sdB+dM & p+g & 255.63 (30)  & 3.01  & 3712.369               & 29.6 & 5.42 & 0.1258 &   0.1258 & 8,27\\
E246387816 & EQ\,Psc & sdB+dM & g+p & 233  & 2.21  & 495.8                      & 28.69 (5) & 5.64 (1) & 0.80083 (1) & 9.4 & 28 \\
E228755638 & HW\,Vir & sdB+dM & g+p  & --  & 0.100  & 309.2     & 28.07 (5) & 5.51 (1) & 0.117 & 0.117 & 29 \\
K11179657 &  & sdB+dM & g+p     & 259.6 (14)  & 1.66  & 186.5                   & 26 & 5.14 & 0.394 & 7.2 &  30,31\\
E246023959 & PHL\,457 & sdB+dM & g+p    & 259 (2) & 1.96  & 265.5       & 26.69 (6) & 5.31 (1) & 0.3128903 (4) & 2.5/4.6 & 28\\

K2991403 &  & sdB+dM & g & 268.52 (74)  & 1.07  & 334.8                 & 27.3 & 5.43 & 0.443 & 10.46 &  30,31\\
K2438324 & NGC\,6791\,B4  & sdB+dM & g & 236.2 (2.1)  & 1.40  & 216.3   & 27.10 (82) & 5.69 (10) & 0.398 & 9.21 (18) & 10,31\\ \hline
%E211623711$^{\dagger}$& UVO\,0825+15 &  He-sdB & -- &38.90 (27) & 5.97 (11) &  -- & -- & 34\\ \hline
\end{tabular}
\end{table}
\end{landscape}

\end{document}